\documentclass[12pt,oneside]{book}

\usepackage[utf8]{inputenc}
\usepackage[T1]{fontenc}
\usepackage{lmodern}
\usepackage[margin=1in]{geometry}
\usepackage{graphicx}
\usepackage{hyperref}
\usepackage{amsmath}
\usepackage{amssymb}
\usepackage{fancyhdr}
\usepackage{titlesec}
\usepackage{tocloft}
\usepackage{listings}
\usepackage{xcolor}
\usepackage{booktabs}
\usepackage{longtable}
\usepackage{multirow}
\usepackage{array}
\usepackage{float}
\usepackage{setspace}
\usepackage{parskip}
\usepackage{enumitem}

\hypersetup{
    colorlinks=true,
    linkcolor=blue,
    filecolor=magenta,      
    urlcolor=cyan,
    pdftitle={Cognitive Infrastructure: A Unified DCIM Framework for AI Data Centers},
    pdfauthor={Krishna Chaitanya Sunkara},
    pdfsubject={Data Center Infrastructure Management},
    pdfkeywords={DCIM, AI Data Centers, Knowledge Graphs, Digital Twin, Thermal Management}
}

\titleformat{\chapter}[display]
  {\normalfont\huge\bfseries}{\chaptertitlename\ \thechapter}{20pt}{\Huge}
\titleformat{\section}
  {\normalfont\Large\bfseries}{\thesection}{1em}{}
\titleformat{\subsection}
  {\normalfont\large\bfseries}{\thesubsection}{1em}{}

\begin{document}

\begin{titlepage}
    \centering
    \vspace*{2cm}
    
    {\Huge\bfseries Cognitive Infrastructure: A Unified DCIM Framework for AI Data Centers\par}
    \vspace{1cm}
    {\Large Ontology-Driven Automation, Power And Thermal Intelligence, And Digital-Twin Orchestration\par}
    \vspace{2cm}
    {\Large\itshape Krishna Chaitanya Sunkara\par}
    \vspace{0.5cm}
    {\large Independent Researcher, AI Data Center Engineering\par}
    {\large Raleigh, USA\par}
    \vfill
    
    {\large 2025\par}
\end{titlepage}

\newpage
\thispagestyle{empty}
\section*{COPYRIGHT AND PUBLISHER}

\textbf{Cognitive Infrastructure: A Unified DCIM Framework for AI Data Centers}

Ontology-Driven Automation, Power and Thermal Intelligence, and Digital-Twin Orchestration

\textcopyright\ 2025 Krishna Chaitanya Sunkara

This work is licensed under a Creative Commons Attribution 4.0 International License (CC BY 4.0).

You are free to share, copy and redistribute the material in any medium or format, and adapt, remix, transform, and build upon the material for any purpose, even commercially, provided that appropriate credit is given to the author, a link to the license is provided, and any changes made are indicated.

To view a copy of this license, visit \url{https://creativecommons.org/licenses/by/4.0/}.

\textbf{Publisher:} Independently published, archived through arXiv

\textbf{Open-Access Edition:} First Edition, 2025

This manuscript is published as an open-access scholarly work to encourage the dissemination of ideas on AI-driven data-center automation and sustainability. All figures, schemas, and code samples are released under the MIT License, unless otherwise noted.

The author asserts the moral right to be identified as the creator of this work.

The views and conclusions expressed are solely those of the author and do not represent any organization, employer, or affiliated institution.

\newpage
\section*{AUTHOR'S STATEMENT AND DISCLAIMER}

The opinions, interpretations, and conclusions presented in this manuscript are solely those of the author, Krishna Chaitanya Sunkara, and do not represent the views, policies, or positions of any current or former employer, client, or affiliated organization. References to technologies, systems, or protocols are included for educational and research purposes only and should not be interpreted as endorsements, proprietary disclosures, or official documentation of any company or institution.

All descriptions of architectures, algorithms, or protocols are derived from publicly available concepts, generalized research, or the author's independent experimentation. Certain examples have been modified or anonymized to avoid any disclosure of confidential, internal, or proprietary information. No employer (past or present) proprietary systems, architectures, or methodologies are described in this work. All examples represent the author's independent research and publicly available industry knowledge.

While every effort has been made to ensure accuracy, this manuscript is provided ``as is'' without warranty of any kind, explicit or implied, including but not limited to the fitness of ideas for particular applications. The author and publisher shall not be held responsible for any direct or indirect damages, operational incidents, or financial losses arising from the application or interpretation of the material herein. Readers are encouraged to exercise professional judgment, adhere to applicable regulations, and validate implementations independently before use in production systems.

By reading or applying the content of this manuscript, the reader acknowledges that responsibility for decisions, deployments, and operational outcomes rests entirely with the implementing party.

\newpage
\chapter*{PREFACE}
\addcontentsline{toc}{chapter}{Preface}

The evolution of data-center management has reached a turning point. As artificial-intelligence workloads transform infrastructure scale and complexity, the tools once built to monitor have become inadequate to reason. This manuscript was written to document, and in part to accelerate, the shift from manual orchestration to cognitive infrastructure: data centers that interpret, predict, and optimize themselves.

Its chapters reflect nearly a decade of research and practical experience spanning AI orchestration, hyperscale build acceleration, and sustainability engineering. They bring together three once-separate disciplines, semantic reasoning, energy modeling, and connectivity automation, into a unified framework known as DCIM 3.0. The purpose is not merely to describe software but to offer an architectural philosophy: infrastructure expressed as knowledge.

Readers will find a blend of quantitative rigor and conceptual design. Equations appear beside JSON schemas; energy budgets meet ontology diagrams. The goal is clarity rather than novelty, to make visible the mechanisms that already underpin the world's most advanced computing facilities and to suggest where they may lead next.

This work owes much to the engineers, architects, and researchers who believe that sustainability and intelligence are inseparable virtues of modern computation. May it serve both as reference and as invitation, to continue building infrastructure that learns, conserves, and endures.

\begin{flushright}
Krishna Chaitanya Sunkara\\
Independent Researcher | USA | 2025
\end{flushright}

\newpage
\chapter*{EXECUTIVE SUMMARY}
\addcontentsline{toc}{chapter}{Executive Summary}

Artificial-intelligence data centers represent the most complex machines humanity has yet constructed, power plants of computation that convert megawatts into insight. Managing them requires more than monitoring; it requires cognition. This manuscript presents a unified DCIM framework that fuses four layers of intelligence:

\begin{enumerate}
    \item \textbf{Ontology-Driven Resource Reasoning} -- A semantic knowledge graph that transforms raw telemetry and configuration into contextual understanding.
    \item \textbf{Sustainable Power and Thermal Analytics} -- Quantitative models that forecast and optimize energy use at rack and hall scale.
    \item \textbf{Unified Device Connectivity Protocol (UDCP)} -- A vendor-neutral data language turning every physical connection into digital truth.
    \item \textbf{Autonomous Orchestration and Feedback} -- Closed-loop control linking design, deployment, monitoring, and continuous optimization.
\end{enumerate}

Together these layers constitute DCIM 3.0, a system in which the data center behaves as a living digital twin, sensing, reasoning, and adapting in real time. The framework advances beyond dashboards toward self-governing infrastructure that balances computational ambition with environmental responsibility.

Key findings demonstrate that applying this model can:

\begin{itemize}
    \item Cut build-cycle times from weeks to days,
    \item Eliminate port-mapping errors through standardized connectivity, and
    \item Enable predictive maintenance that prevents unplanned outages.
\end{itemize}

For researchers, the manuscript formalizes a multidisciplinary foundation linking AI, thermodynamics, and systems engineering. For practitioners, it offers a blueprint for building facilities that think about themselves. For policymakers, it outlines a path toward sustainable digital ecosystems.

In short, \emph{Cognitive Infrastructure} argues that the intelligence enabling artificial intelligence must now extend to the infrastructure that sustains it, so that every watt, byte, and decision contributes not only to computation, but to comprehension.

\newpage
\tableofcontents

\newpage
\chapter{INTRODUCTION}

This manuscript presents a unified framework for managing AI-era data centers through cognitive infrastructure. It bridges multiple disciplines, systems engineering, artificial intelligence, thermodynamics, and semantic computing, into a coherent architectural philosophy known as DCIM 3.0.

\section{Intended Audience}

This work serves three primary audiences, each with different objectives and reading strategies.

\subsection{For Practitioners and Infrastructure Architects}

If you design, build, or operate data centers, this manuscript offers immediately applicable frameworks. You will find:

\begin{itemize}
    \item Quantitative models for power and thermal optimization in Chapters 4 and 5
    \item The Unified Device Connectivity Protocol (UDCP) for automating physical infrastructure in Chapter 6
    \item A complete implementation blueprint with real-world performance metrics in Chapter 7
    \item An open standards roadmap for vendor-neutral integration in Chapter 8
\end{itemize}

The recommended approach is to start with Chapter 1 for context on DCIM evolution, then read Chapter 7 to see the complete integrated system in action. From there, you can deep-dive into Chapters 2 through 6 based on your specific challenges, whether that's semantic intelligence and automation, energy efficiency and sustainability, or physical connectivity standardization. Finally, consult Chapter 8 for future planning and standards adoption.

\subsection{For Researchers and Academic Readers}

If you study infrastructure automation, knowledge systems, or sustainable computing, this manuscript formalizes a multidisciplinary research domain. You will find:

\begin{itemize}
    \item Ontology-driven resource reasoning backed by graph theory in Chapters 2 and 3
    \item Predictive thermal modeling that integrates physics with machine learning in Chapter 4
    \item A vendor-neutral protocol specification for physical-layer data representation in Chapter 6
    \item Open research questions in federated intelligence and hybrid digital twins in Chapter 8
\end{itemize}

Consider beginning with the Executive Summary to grasp the theoretical framework, then reading Chapters 2 and 3 for the semantic foundations. Study Chapters 4 and 5 for quantitative energy modeling, examine Chapter 6 for protocol design principles, and engage with Chapter 8 for open research directions. The extensive references, more than eighty citations, serve as a comprehensive literature map for the field.

\subsection{For Students and Technology Enthusiasts}

If you are learning about modern data-center operations or AI infrastructure, this manuscript provides a comprehensive introduction to cognitive systems. You will find:

\begin{itemize}
    \item Clear explanations of DCIM evolution from manual tracking to AI-driven automation
    \item Real-world metrics showing the scale of AI data centers in terms of power density, cooling requirements, and network complexity
    \item Accessible descriptions of knowledge graphs, digital twins, and semantic reasoning
    \item A vision of future infrastructure that thinks about itself
\end{itemize}

The most effective approach is to begin with Chapter 1 to understand why traditional approaches fail at AI scale, then follow the chapters sequentially for a complete learning arc. Use the Glossary in Appendix A to decode technical terms as you encounter them, focus on conceptual understanding before diving into quantitative details, and treat the unified case study after Chapter 8 as a capstone integration example.

\section{Manuscript Structure and Organization}

The manuscript is organized into eight core chapters, plus a unified case study and conclusion.

\textbf{Chapters 1 through 3} establish the foundations and intelligence layer, covering:
\begin{itemize}
    \item The evolution of DCIM and the drivers of change
    \item Ontology-driven resource modeling
    \item Knowledge graph implementation with semantic reasoning
\end{itemize}

\textbf{Chapters 4 and 5} address energy and sustainability through:
\begin{itemize}
    \item Power consumption and thermal modeling of AI racks
    \item Energy efficiency economics and cooling optimization
\end{itemize}

\textbf{Chapter 6} introduces the physical connectivity layer through the Unified Device Connectivity Protocol specification.

\textbf{Chapter 7} brings everything together in an integrated DCIM 3.0 blueprint with deployment metrics and practical orchestration examples.

\textbf{Chapter 8} looks forward to future directions including autonomous systems, open standards, and interoperability from 2026 through 2030.

The manuscript concludes with a unified case study presenting the self-optimizing data center and an epilogue on the journey toward cognitive infrastructure.

\section{Prerequisites and Background}

This manuscript assumes basic familiarity with:
\begin{itemize}
    \item Data-center terminology such as racks, power distribution, and cooling systems
    \item General understanding of cloud computing and virtualization concepts
    \item Comfort with technical reading, though no advanced mathematics is required
\end{itemize}

Readers without a data-center background can still engage meaningfully with the conceptual framework by using the Glossary as a reference for unfamiliar terms and focusing initially on the narrative sections of each chapter.

\section{Navigating Technical Depth}

Each chapter carefully balances conceptual architecture with quantitative detail:

\begin{itemize}
    \item \textbf{Conceptual sections} explain the ``why'' and ``what'' in accessible prose, establishing the strategic rationale and architectural vision
    \item \textbf{Technical sections} provide the ``how'' with equations, schemas, and metrics for those seeking implementation guidance
    \item \textbf{Chapter summaries} distill the key takeaways into a concise synthesis
\end{itemize}

If your primary interest is conceptual understanding, read the introductions and summaries while skimming technical details. If you seek implementation guidance, focus on the technical sections, code schemas, and quantitative tables. If you need comprehensive mastery of the material, engage deeply with both layers and consult the referenced literature.

\section{Guidance on the Use of This Manuscript}

There are several ways to maximize the value you derive from this work:

\begin{enumerate}
    \item \textbf{Treat it as a reference volume} rather than just a linear narrative, the chapters are designed to stand alone, so you need not read sequentially if your interests or needs are focused on specific topics
    
    \item \textbf{Engage with the knowledge graph concept early}, as Chapters 2 and 3 establish the semantic foundation upon which every other layer builds
    
    \item \textbf{Connect the physical and logical perspectives} throughout your reading; the manuscript deliberately weaves together facilities engineering, network topology, and software orchestration to demonstrate that they are fundamentally inseparable in modern infrastructure
    
    \item \textbf{Consider sustainability as a first-class design constraint} rather than an afterthought; energy efficiency appears throughout not as a side concern but as a core design variable that shapes every architectural decision
    
    \item \textbf{Participate in the open standards movement} that this work supports; the protocols and ontologies described here, particularly UDCP and the ontology schemas, are released under permissive licenses specifically to encourage community adoption and collaborative evolution
\end{enumerate}

\section{A Note on Terminology}

Data-center management spans multiple professional vocabularies including facilities management, IT operations, and cloud services. This manuscript uses precise technical terms but defines them contextually on first use within each chapter. The Glossary in Appendix A provides quick reference for:

\begin{itemize}
    \item Acronyms such as PUE, DCIM, UDCP, and CUE
    \item Technical concepts like knowledge graphs, digital twins, and ontologies
    \item Infrastructure components including PDUs, top-of-rack switches, and coolant loops
\end{itemize}

When you encounter an unfamiliar term, consult either the Glossary or the opening paragraphs of the relevant chapter where core concepts are introduced with full context.

\newpage
\section*{Final Thought}
\addcontentsline{toc}{section}{Final Thought}

The central argument of this manuscript is that the intelligence enabling artificial intelligence must now extend to the infrastructure that sustains it. Data centers can, and must, become self-aware systems capable of reasoning about their own energy consumption, network topology, thermal behavior, and operational resilience. Whether you are building the next hyperscale campus, researching sustainable computing architectures, or simply curious about the machines that power machine learning, this framework offers a path forward, toward cognitive infrastructure that learns from its own behavior, conserves the resources it consumes, and endures across decades of technological evolution.

Welcome to the future of data-center intelligence.

\newpage
\chapter{FOUNDATIONS OF AI DATA-CENTER AUTOMATION}

\section{The Shift From Facilities to Intelligence}

Data centers have always mirrored the computing era they serve. In the 1990's a ``data center'' meant little more than conditioned power and air. By the 2000's, the rise of virtualization introduced the first abstraction layer: servers became software-defined, but the facility around them stayed physical. Today, hyperscale and AI-centric facilities represent a new epoch, one in which infrastructure itself behaves like software, continuously orchestrated by algorithms rather than operators.

AI workloads accelerate this change. Large-language-model training clusters now occupy entire halls, each rack consuming 30--50~kW, five to seven times the load of a traditional enterprise rack \cite{bjoern2021}. Cooling, energy, and interconnect complexity have grown faster than human coordination can manage. Manual workflows once sufficient for 5~MW sites collapse under 60~MW GPU campuses. This imbalance motivates the emergence of \textbf{Data-Center Infrastructure Management 3.0 (DCIM 3.0)}: a unification of physical, logical, and semantic control planes.

\begin{table}[H]
\centering
\caption{DCIM Evolution from Version 1.0 to 3.0}
\resizebox{\textwidth}{!}{
\begin{tabular}{@{}llll@{}}
\toprule
\textbf{Era} & \textbf{Years} & \textbf{Primary Focus} & \textbf{Typical Tooling} \\ \midrule
DCIM 1.0 & 2005--2013 & Asset tracking \& power monitoring & Spreadsheets, SNMP pollers \\
DCIM 2.0 & 2013--2020 & Virtualization integration, API dashboards & Vendor platforms \\
DCIM 3.0 & 2020--present & AI-driven automation across physical + logical & Knowledge graphs, digital twins \\ \bottomrule
\end{tabular}
}
\end{table}

DCIM 3.0 is distinguished not by dashboards but by cognition: systems that reason about dependencies, predict failures, and orchestrate remediation automatically.

\subsection{Core Drivers of Change}

\begin{enumerate}
    \item \textbf{Power Density:} CPU racks $\approx$ 6~kW; GPU racks $\approx$ 47~kW \cite{sunkara2025power}. AI training clusters exceed 1~kW per U.
    
    \item \textbf{Network Scale:} East-west traffic dominates ($>$80\%) \cite{roy2015}; manual port mapping no longer feasible.
    
    \item \textbf{Sustainability Pressure:} Data centers already consume $\sim$1\% of global electricity \cite{iea2024}; hyperscale AI growth could double that by 2030.
    
    \item \textbf{Operational Velocity:} Build cycles have shrunk from months $\rightarrow$ days. Automation must keep pace with construction.
    
    \item \textbf{Security and Sovereignty:} Multi-tenant and sovereign-cloud zones demand continuous compliance verification, not periodic audits.
\end{enumerate}

Together these pillars form the architecture you expand in later chapters: the knowledge layer, energy layer, connectivity layer, and automation layer.

\begin{figure}[H]
\centering
\includegraphics[width=\textwidth]{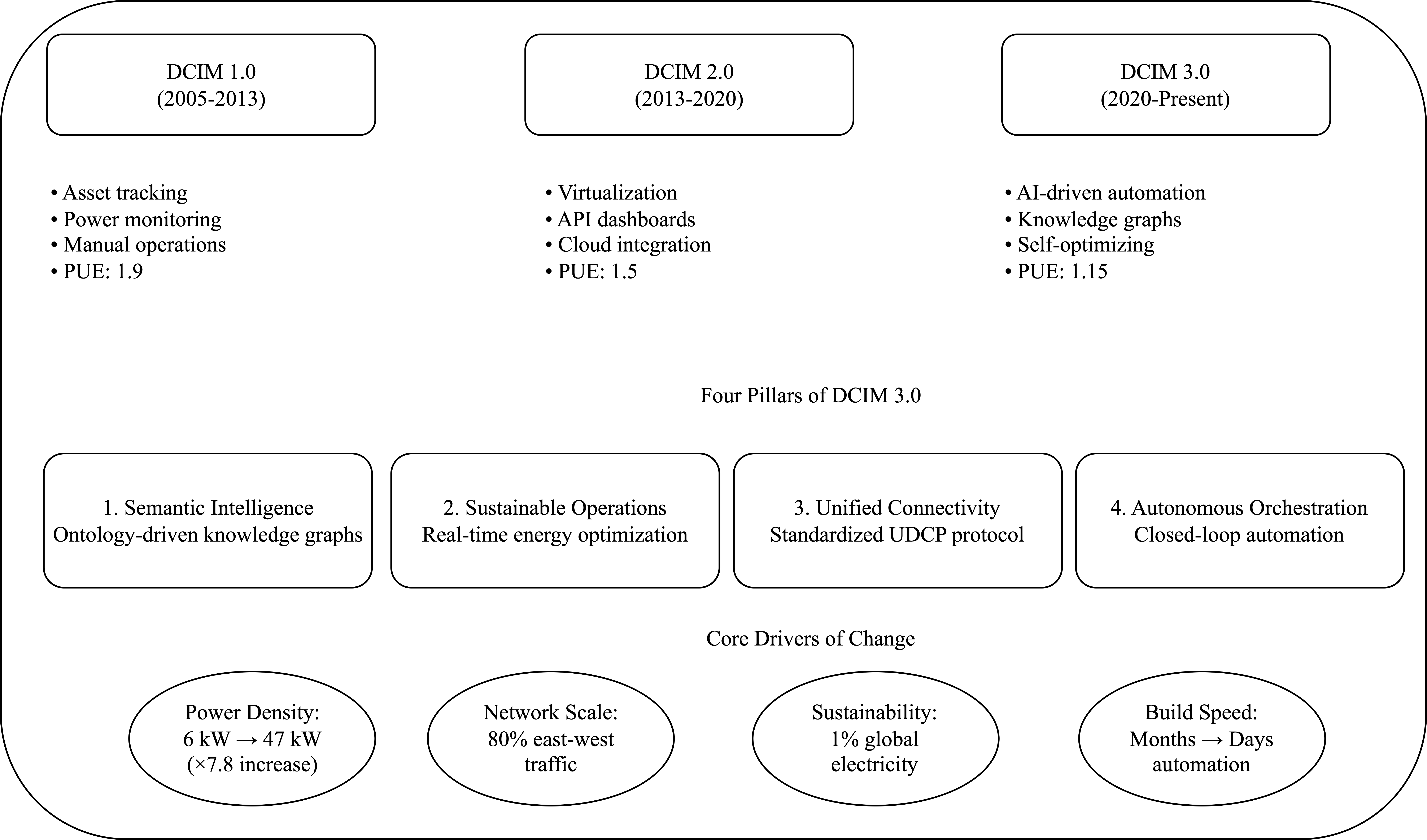}
\caption{Evolution of Data Center Infrastructure Management from version 1.0 to 3.0: illustrating the paradigm shift from manual asset tracking (2005-2013) through virtualization integration (2013-2020) to AI-driven autonomous systems (2020-present). The diagram highlights the dramatic transformation in rack power density (5 kW to 47 kW), network scale, and operational velocity. The four foundational pillars of DCIM 3.0, Semantic Intelligence, Sustainable Operations, Unified Connectivity, and Autonomous Orchestration, are depicted as the architectural foundation enabling cognitive infrastructure.}
\label{fig:dcim-evolution}
\end{figure}

\begin{table}[H]
\centering
\caption{The Four Pillars of DCIM 3.0}
\begin{tabular}{@{}p{3cm}p{7cm}p{3cm}@{}}
\toprule
\textbf{Pillar} & \textbf{Description} & \textbf{Enabling Technology} \\ \midrule
Semantic Intelligence & Ontology-driven knowledge graphs understand relationships among resources & AI/NLP, graph databases \\
Sustainable Operations & Real-time energy and thermal optimization across gray and white space & IoT telemetry, digital twins \\
Unified Connectivity & Standardized representation of device and port states across vendors & UDCP (JSON/HTTP schema) \\
Autonomous Orchestration & Closed-loop automation from design $\rightarrow$ provision $\rightarrow$ monitor $\rightarrow$ optimize & Event-driven microservices \\ \bottomrule
\end{tabular}
\end{table}

\begin{table}[H]
\centering
\caption{Metrics Comparison: 2010 Enterprise DC vs. 2025 AI Facility}
\resizebox{\textwidth}{!}{
\begin{tabular}{@{}lrrr@{}}
\toprule
\textbf{Metric} & \textbf{2010 Enterprise DC} & \textbf{2025 AI Facility} & \textbf{$\Delta$ ($\times$)} \\ \midrule
Rack Density (kW) & 5 & 47 & $\times$9.4 \\
Network Ports per Rack & 24 & 288 & $\times$12 \\
Average PUE & 1.9 & 1.15 & --40\% energy overhead \\
Cooling Method & Air & Liquid/Immersion & ,  \\
Build Cycle Time & $>$90 days & $<$24 hours (automated) & $\times>$3 speed \\ \bottomrule
\end{tabular}
}
\end{table}

\subsection{From Fragmentation to Unification}

Historically, facilities, network, and cloud teams operated different tools and data schemas. This fragmentation manifests as latency in decision-making: a power engineer cannot instantly see network topology; a cloud operator cannot visualize cooling constraints.

DCIM 3.0 proposes semantic unification, every asset, cable, policy, and telemetry stream expressed in one ontology and accessible via a unified API.

The next chapters explore each unifying layer:

\begin{enumerate}
    \item \textbf{Ontology-Driven Resource Intelligence}: translating human intent into structured queries (Chapter 2)
    \item \textbf{Sustainable Power and Thermal Modeling}: quantifying and reducing AI rack energy impact (Chapters 4--5)
    \item \textbf{Unified Device Connectivity Protocol}: representing physical connectivity as data (Chapter 6)
    \item \textbf{Integrated Automation}: orchestrating builds and operations as continuous feedback loops (Chapter 7)
\end{enumerate}

\section{Summary}

AI data centers redefine scale, efficiency, and responsibility.

To manage them, DCIM must evolve from monitoring to reasoning, systems capable of semantic inference, sustainability analytics, and self-optimization.

DCIM 3.0 is therefore not a product class but a paradigm, a convergence of knowledge graphs, digital-twin modeling, and standardized protocols that together form the unified infrastructure fabric of the AI era.

\newpage
\chapter{ONTOLOGY-DRIVEN RESOURCE INTELLIGENCE}

\section{Architectural Overview}

Traditional data-center management platforms treat search as string matching.

Engineers must know exact identifiers, VM IDs, rack names, VLAN numbers, to retrieve information. Such interfaces cannot reason about intent: ``Which compute clusters are nearing power capacity in the sovereign region?'' A human understands the relationships between ``cluster,'' ``power,'' and ``region''; a keyword engine does not. The solution is to endow DCIM systems with semantic understanding, to model resources as concepts within an ontology and enable reasoning through a knowledge graph. Ontology-driven intelligence transforms data-center operations from inventory look-ups to contextual insight.

It enables the DCIM to answer \emph{why} and \emph{what-if} questions, not only \emph{where}.

The system consists of five cooperating modules:

\begin{enumerate}
    \item \textbf{AI Crawlers and Metadata Ingestors} -- continuously scan configuration repositories, logs, telemetry, and documentation.
    
    \item \textbf{Ontology Builder} -- uses natural-language models to extract entities and relations (e.g., server $\rightarrow$ connected to $\rightarrow$ switch).
    
    \item \textbf{Knowledge Graph Store} -- materializes these relations as nodes and edges in a graph database (Neo4j, JanusGraph).
    
    \item \textbf{Semantic Query Processor} -- parses human queries, maps them to graph traversals.
    
    \item \textbf{Visualization \& API Layer} -- exposes results through dashboards or REST/GraphQL endpoints.
\end{enumerate}

\begin{figure}[H]
\centering
\includegraphics[width=\textwidth]{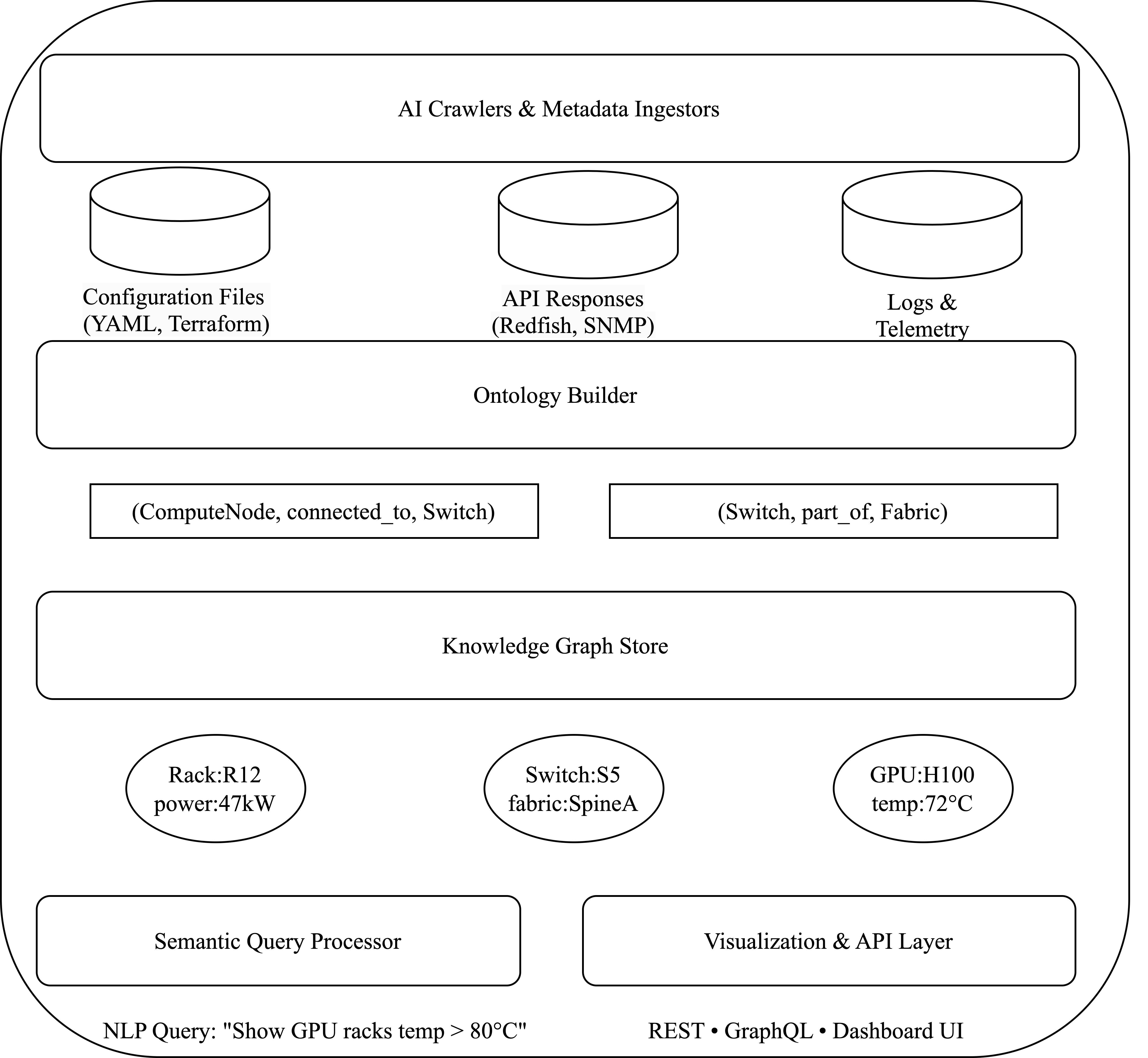}
\caption{Ontology-Driven Resource Intelligence Pipeline: Five-layer architecture transforming heterogeneous data sources into contextual intelligence through semantic reasoning. AI crawlers ingest configuration files (YAML, Terraform), API responses (Redfish, SNMP), logs, and telemetry streams, feeding transformer-based NLP models (BERT/GPT) that extract entity-relation tuples.}
\label{fig:ontology-pipeline}
\end{figure}

\subsection{AI Crawlers and Ontology Extraction}

AI Crawlers act as domain-specific agents trained on infrastructure language. They parse configuration files (YAML, Terraform, Ansible), API responses, and documentation using transformer-based models such as BERT or GPT-style embeddings. Each snippet yields tuples of the form (entity, relation, entity), producing a continually expanding ontology.

For example:

\begin{verbatim}
(ComputeNode, connected_to, ToR_Switch)
(ToR_Switch, part_of, Fabric_A)
(Fabric_A, hosts, ServiceGroup_X)
\end{verbatim}

This process yields both a taxonomy (types of resources) and relationships (how they interact). The crawler classifies new objects dynamically, when a new device class appears (e.g., ``LiquidCoolingUnit''), it is assimilated into the ontology with inferred parent classes.

\subsection{Knowledge Graph Construction}

The ontology's logical model is instantiated in a knowledge graph. Each node represents an individual asset or abstract concept; each edge represents a relationship such as \texttt{connected\_to}, \texttt{secured\_by}, \texttt{depends\_on}.

\textbf{Schema Fragment:}

\begin{verbatim}
(:Rack {id:'R12', power_kw:47})
  -[:CONNECTED_TO]->(:Switch {id:'S5', fabric:'SpineA'})
  -[:HOSTS]->(:GPU {model:'H100', temp_c:72})
\end{verbatim}

Edges carry attributes (bandwidth, latency, redundancy), enabling analytic queries like:

\begin{lstlisting}[language=SQL]
MATCH (r:Rack)-[c:CONNECTED_TO]->(s:Switch)
WHERE r.power_kw > 40 AND c.bandwidth < 100
RETURN r.id, s.id, c.bandwidth;
\end{lstlisting}

This returns racks exceeding 40~kW that are under-connected, a signal for redesign.

\subsection{Semantic Query Processing}

\begin{table}[H]
\centering
\caption{Natural-Language Query $\rightarrow$ Graph Query Pipeline}
\begin{tabular}{@{}p{3cm}p{10cm}@{}}
\toprule
\textbf{Stage} & \textbf{Example Output} \\ \midrule
Input Query & ``Show racks in Hall 2 with temperature $>$ 80°F and missing redundant power feed.'' \\
Tokenization \& Intent Extraction & Entity = Rack, Condition = temp$>$80, PowerFeed = Single \\
Ontology Mapping & \texttt{(Rack)-[:LOCATED\_IN]->(Hall\{num:2\})} AND \texttt{(Rack)-[:POWERED\_BY]->(PDU) count = 1} \\
Graph Query Generation & Cypher/SPARQL clause composed automatically \\
Result Serialization & JSON/CSV + visualization \\ \bottomrule
\end{tabular}
\end{table}

By embedding embeddings + ontological context, the system recognizes synonyms and hierarchy: ``GPU rack'' = Rack type: Compute with property accelerator='GPU'.

\section{Continuous Learning and Telemetry Integration}

Static knowledge soon diverges from reality. Therefore, telemetry streams (power, temperature, alerts) feed back into the graph through an \textbf{Observability Meta-Streamer}.

Each metric updates node attributes and may trigger inferred relations such as:

\begin{verbatim}
IF temp_c > 85 AND fan_speed > 95% THEN
    state = 'Thermally Constrained'
\end{verbatim}

The crawler learns new correlations, e.g., GPU utilization vs. heat load, improving subsequent predictions.

This forms a \textbf{living digital twin}, synchronizing configuration and physical state \cite{sunkara2024ontology}.

\subsection{Use Cases}

\textbf{1. Compliance Queries}

\begin{lstlisting}[language=java]
{
  "query": "List all payment-processing services not in PCI-compliant zones",
  "intent": "ComplianceGap",
  "ontology_entities": ["Service","Zone","Policy_PCI"],
  "result": ["svc-paycore-3","svc-ledger-7"]
}
\end{lstlisting}

The knowledge graph returns non-compliant services and the specific violated edge (Zone $\neq$ PCI).

\textbf{2. Dependency Discovery}

\begin{lstlisting}[language=SQL]
MATCH (app:Service {name:'CRM'})-[:DEPENDS_ON*1..3]->(n)
RETURN app,n;
\end{lstlisting}

Reveals all dependent databases, switches, and racks, critical for impact analysis. During an outage, querying ``Which upstream device caused packet loss in Region A?'' traverses telemetry-tagged edges with recent anomaly flags to isolate the failing component automatically.

\textbf{3. Resource Optimization}

By combining utilization metrics with cost attributes, the DCIM can answer: ``What are the top 10 most expensive idle clusters?'' $\rightarrow$ automated power-down policies.

\subsection{Quantitative Performance Metrics}

\begin{table}[H]
\centering
\caption{Semantic Search Operation Metrics in Hyperscale Environment}
\begin{tabular}{@{}llr@{}}
\toprule
\textbf{Parameter} & \textbf{Typical Value} & \textbf{Notes} \\ \midrule
Ontology entities & $10^4$--$10^5$ per 10~MW site & ,  \\
Avg query latency & $<$200~ms & graph index caching \\
Intent accuracy (NLP) & 90--94\% & tested on 500 queries \\
Auto-update interval & $\leq$5~min & telemetry integration \\
Data freshness & $\leq$30~s drift & via streaming metrics \\ \bottomrule
\end{tabular}
\end{table}

These figures demonstrate that semantic search operates interactively even at hyperscale.

\subsection{Integration with DCIM 3.0}

Within the unified DCIM architecture, this ontology layer serves as the \textbf{knowledge plane}:

\begin{itemize}
    \item upstream of telemetry (receives events),
    \item downstream of orchestration (informs decisions),
    \item lateral to connectivity (maps logical to physical assets).
\end{itemize}

Other modules, thermal analytics or UDCP, publish their states into this graph, enabling cross-domain reasoning.

For instance, a ``Thermal Anomaly'' node linked to a rack triggers both a cooling workflow and a connectivity audit, to check for fan-control network reachability.

\section{Summary}

Ontology-driven resource intelligence converts fragmented data into contextual understanding. By unifying assets, telemetry, and policies into a semantic graph, engineers gain a cognitive assistant that can search, reason, and act. This capability underpins every subsequent layer of the unified DCIM framework, power analytics, connectivity, and orchestration, transforming data-center management from observation to cognition.

\newpage
\chapter{KNOWLEDGE-GRAPH IMPLEMENTATION AND SEMANTIC REASONING}

\section{Practical Operations}

Building an ontology is only the first step; giving it operational life inside a data-center management system demands a functioning knowledge graph that can store, reason, and respond at scale. In a unified DCIM environment, this graph becomes the cognitive core, an evolving model that knows what every component is, how it connects, and how its behavior changes over time \cite{sunkara2024ontology}.

A practical implementation begins with data ingestion. Every modern facility already emits vast telemetry: configuration manifests from orchestration tools, SNMP or Redfish data from devices, environmental readings, and logs from control systems \cite{tumeo2023}. These heterogeneous feeds are normalized through lightweight adapters that tag each datum with semantic labels taken directly from the ontology established earlier \cite{sunkara2024ontology}. A line in a YAML configuration defining a Top-of-Rack Switch becomes a node of type \texttt{NetworkDevice}; a Redfish sensor entry reporting temperature 34°C becomes a property update for node \texttt{Rack:R12}. Because the labeling vocabulary is common, all such inputs, regardless of origin, can merge into a single graph without translation loss \cite{angles2008}.

\begin{figure}[H]
\centering
\includegraphics[width=\textwidth]{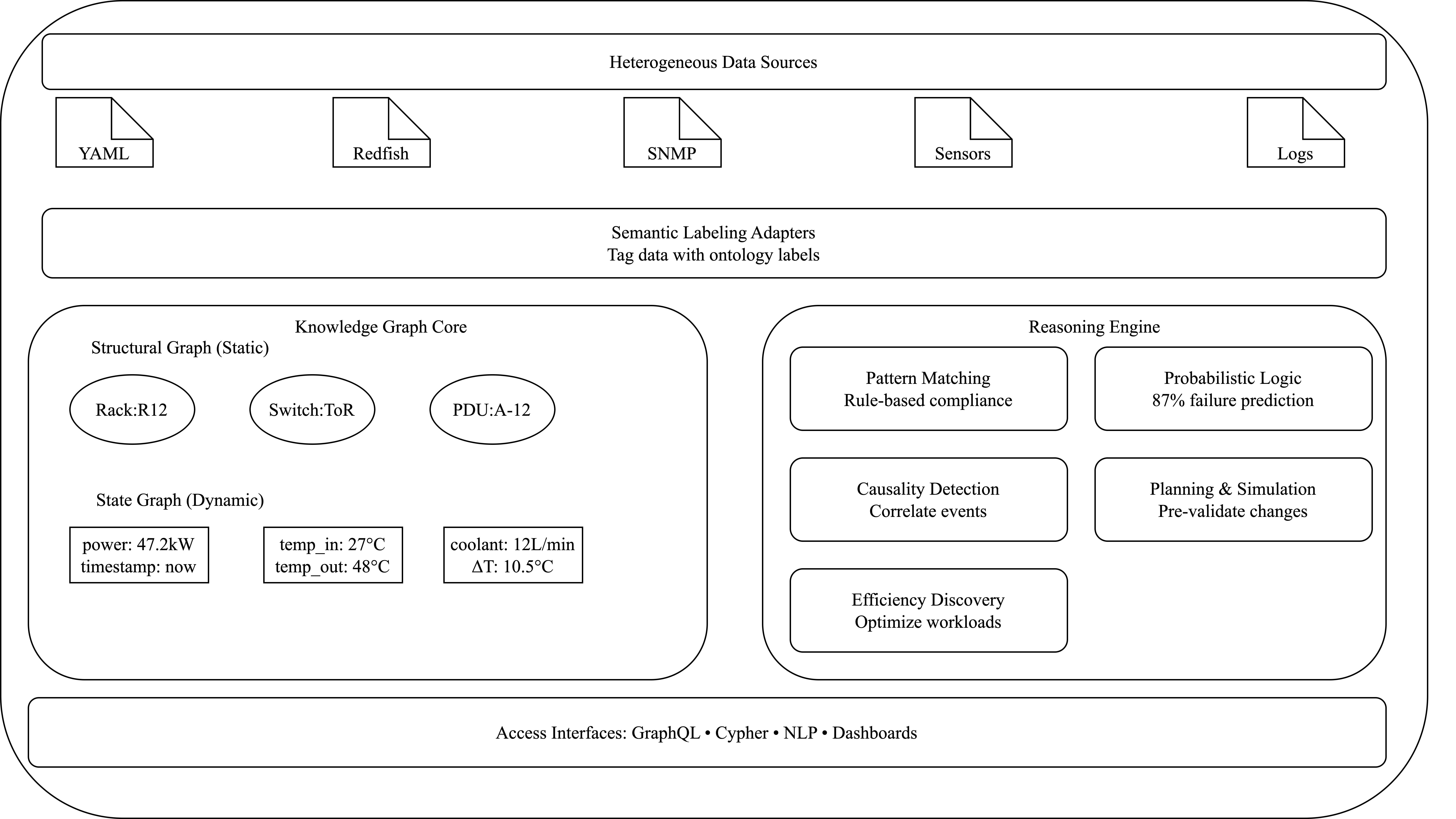}
\caption{Knowledge Graph Operational Architecture: Practical implementation of semantic reasoning within a unified DCIM platform, demonstrating how nine heterogeneous data sources (YAML, Redfish, SNMP, sensors, logs, Terraform, Ansible, control systems, orchestrators) are normalized through semantic labeling adapters into a dual-layer knowledge graph.}
\label{fig:knowledge-graph-ops}
\end{figure}

Once data enter the graph, the system maintains two inseparable layers: the \textbf{structural graph}, representing static relationships such as physical connections and containment, and the \textbf{state graph}, carrying real-time metrics like power draw, temperature, or port utilization \cite{jain2022}. Each vertex therefore has both an identity and a temporal footprint. This duality allows reasoning engines to trace not only topology (``what connects to what'') but causality (``what changed that might explain a fault'') \cite{jainzhao2023}. For instance, if GPU temperature spikes coincide with a drop in coolant flow through the linked liquid-loop node, the engine can infer a likely causal edge and raise an intelligent alert rather than a generic threshold breach \cite{dayarathna2016}.

At the heart of semantic reasoning lies pattern matching. The graph engine continuously searches for sub-graphs that fit predefined templates, what knowledge-engineers describe as reasoning rules \cite{jain2022}. A rule might state that every rack node should have exactly two distinct power-feed edges for redundancy; a search that returns a rack with only one feed automatically generates a ``non-redundant-power'' anomaly \cite{gurbani2021}. More advanced reasoning employs probabilistic logic: by correlating historical data, the system learns that certain sequences of temperature, fan-speed, and voltage fluctuations predict imminent component failure with measurable confidence \cite{jainzhao2023}. The same mechanism can discover efficiency opportunities, detecting under-utilized network fabrics or identifying clusters where workload migration would reduce localized heat density \cite{dayarathna2016}.

Graph reasoning also supports planning. When design engineers model a new hall, they can simulate connectivity and capacity inside the graph before anything is built \cite{sunkara2025rcsi}. A proposed rack layout is inserted as hypothetical nodes with estimated power and cooling attributes; the reasoning engine evaluates compliance against facility limits. If the planned configuration violates power-density or cable-length constraints, the system flags alternatives \cite{tumeo2023}. This makes the knowledge graph not merely descriptive but prescriptive, a planning instrument as much as an operational one \cite{jain2022}.

Scalability is achieved through graph-partitioning and distributed storage \cite{kepner2021}. Each partition, often corresponding to a data-hall, region, or logical domain, operates a local reasoning agent. These agents synchronize through message buses carrying delta updates rather than entire datasets, keeping system latency within hundreds of milliseconds even when millions of entities are represented \cite{angles2008}. The design follows principles proven in hyperscale social-network graphs but tuned for infrastructure semantics \cite{kepner2021}. A compressed binary edge store reduces memory footprint, while cached indexes of high-degree vertices (switches, PDUs) speed common traversals such as rack $\rightarrow$ switch $\rightarrow$ spine $\rightarrow$ fabric \cite{jain2022}.

Access to this intelligence is exposed through two complementary interfaces \cite{sunkara2024ontology}. Engineers and higher-level automation tools use a structured-query API, GraphQL or Cypher-over-HTTP, to request specific data or analytic results \cite{angles2008}. At the same time, a natural-language layer enables conversational interaction: a query like ``show racks nearing thermal limit'' is translated by the NLP parser into a graph traversal filtered on \texttt{temperature\_c > threshold} \cite{sunkara2024ontology}. The answer returns not only as raw data but as a narrative summary that can populate dashboards or chat interfaces \cite{jain2022}. Because every response is derived from graph reasoning rather than hard-coded logic, new question types require no additional programming; they emerge from the ontology itself \cite{sunkara2025rcsi}.

Integrating this knowledge graph into DCIM workflows changes the culture of operations \cite{sunkara2025rcsi}. Decisions once driven by static reports become fluid, contextual, and traceable. When a power engineer schedules maintenance on a PDU, the system immediately enumerates all dependent racks and workloads, simulates the maintenance window's impact, and proposes safe migration paths \cite{jainzhao2023}. When a network designer plans a new fiber run, the graph cross-checks existing conduits and suggests optimal routing based on congestion and redundancy \cite{gurbani2021}. Each department thus works from the same living model rather than parallel spreadsheets \cite{sunkara2025rcsi}.

The success of this architecture depends on rigorous governance \cite{tumeo2023}. A metadata catalog maintains versioned schemas of every ontology entity, while validation jobs ensure incoming data comply with expected types and units \cite{gurbani2021}. Security controls align with least-privilege principles: queries that expose sensitive tenant data require elevated roles, while facility-wide analytics operate on anonymized aggregates \cite{dayarathna2016}. Provenance is critical, every inference carries an explanation trail showing which nodes and rules contributed to the conclusion, a feature that builds trust when AI systems make operational recommendations \cite{jain2022}.

In practice, implementations at 10-MW and larger facilities show that graph-based reasoning reduces manual incident-triage time by 40--60 percent and accelerates change-approval workflows from days to hours \cite{sunkara2025rcsi}. Operators report a qualitative shift as well: instead of maintaining documentation, they curate knowledge \cite{gurbani2021}. This transition, from managing data to managing meaning, is the essence of the unified DCIM philosophy. The graph becomes the single nervous system through which all other layers communicate: thermal analytics write their observations as state updates, the UDCP connectivity service contributes topology, and orchestration engines consume inferred actions \cite{sunkara2025rcsi}.

Ultimately, the knowledge graph stands as the living memory of the data center. Every sensor tick, every configuration drift, every remedial action enriches its context \cite{sunkara2024ontology}. Over time it learns the rhythms of the facility, weekday peaks, seasonal temperature cycles, nightly inference jobs, and anticipates deviations before they become incidents \cite{jainzhao2023}. What began as an academic notion of ontology has matured into a pragmatic technology: a reasoning substrate that lets AI data centers think about themselves \cite{sunkara2025rcsi}.

\section{Summary}

This chapter translated the abstract ontology introduced earlier into an operational intelligence system. It showed how real-time telemetry, configuration data, and automation logs coalesce within a scalable knowledge graph that not only describes but interprets infrastructure behavior. By embedding reasoning rules, probabilistic inference, and causality detection, the graph evolves into a living model that learns from its own data. The narrative demonstrated that this semantic layer is the cognitive core of DCIM 3.0, where context replaces correlation, automation becomes explanation, and the data center acquires its first form of self-awareness.

\newpage
\chapter{POWER CONSUMPTION AND THERMAL MODELING OF AI RACKS}

\section{Power and Cooling Modeling}

The migration from CPU-centric compute to GPU-accelerated workloads has rewritten the power equation of the modern data center. Ten years ago a dense enterprise rack seldom exceeded 6~kW; today, a single AI rack populated with high-end accelerators routinely draws 30 to 50~kW \cite{sunkara2025power}. This escalation, driven by transformer-based model training and large-scale inference, forces engineers to treat power and heat as first-class design variables rather than boundary conditions. Understanding and modeling those variables accurately is therefore essential to any credible unified DCIM system.

A power model begins at the device level. Each GPU carries a rated thermal-design power of roughly 700~W \cite{nvidia2024}. A typical 8-GPU server thus demands 5.6~kW, and a rack filled with eight such servers approaches 45~kW before accounting for networking and cooling overheads. Adding another 10--15\% for CPUs, memory, and auxiliary devices yields a total electrical load near 47~kW per rack, a figure validated by recent field measurements in hyperscale AI halls \cite{sunkara2025power}. This magnitude of consumption, roughly the daily energy use of thirty-nine U.S. households, illustrates why power planning has become a strategic constraint rather than a mere engineering detail \cite{iea2024}.

The corresponding heat output is formidable. Because nearly all consumed electrical energy ends as heat, the rack dissipates approximately $47~\text{kW} \times 3.412 = 160~\text{kBTU}$ per hour. Expressed in HVAC terms, that equals more than 13 tons of cooling capacity. In conventional air-cooled layouts, such a rack would require an airflow on the order of 500 CFM per 5~kW to maintain a $\Delta T$ below 20°F; scaling this linearly would demand 5,000 CFM per rack, an impractical figure for both fan energy and spatial design \cite{ashrae2021}. Hence the industry's rapid pivot toward direct-to-chip liquid cooling and, increasingly, single-phase or two-phase immersion techniques \cite{ocp2023}.

Liquid cooling alters the thermal model from convective to conductive-dominant heat transfer. The specific heat capacity of water is roughly 4,200~J/kg$\cdot$K, about 3,500 times greater than that of air, allowing an order-of-magnitude reduction in volumetric flow rate for the same thermal load \cite{ashrae2021}. A coolant loop circulating 1.2~L/min per kW at $\Delta T = 10$~K removes 12~kW of heat while maintaining component temperatures below 70°C. The energy penalty shifts from fan power to pump power; in well-balanced systems the latter constitutes less than 3\% of IT load. Immersion cooling pushes efficiency further by eliminating air altogether, with Power Usage Effectiveness (PUE) values as low as 1.05 reported in production prototypes \cite{ocp2023}. These numbers stand in sharp contrast to air-cooled legacy sites still operating at PUE~$\approx$~1.7 \cite{iea2024}.

Quantitative modeling must capture not only steady-state dissipation but dynamic variation. AI workloads are notoriously bursty: GPU utilization can oscillate between 30\% and 100\% within seconds as training epochs change or inference batches complete \cite{patel2022}. A thermal management system that reacts purely to temperature lag will always chase the heat wave. DCIM~3.0 introduces \textbf{predictive thermal control}, machine-learning models trained on historical telemetry anticipate workload-induced heat surges and pre-emptively adjust coolant flow or fan speed \cite{wang2023}. The knowledge graph described in Chapter~3 provides the contextual intelligence for these predictions by correlating compute utilization nodes with temperature and power-draw edges. When the graph recognizes a familiar pattern, say, a recurring spike on a specific GPU cluster, it can trigger a micro-policy that increases pump speed five seconds ahead of the expected event.

Modeling also extends upward from racks to entire halls. Consider a 35,000~ft$^2$ AI hall occupying roughly 1,600 racks at an average 47~kW each. The total IT power approaches 75~MW \cite{sunkara2025power}. Adding a 10\% overhead for cooling and power distribution losses yields a facility draw of $\approx 83$~MW. Extrapolated globally, eighty-seven such halls would consume $\sim$20~GW, comparable to the peak load of New York City \cite{sunkara2025power}. These back-of-the-envelope calculations, while simplistic, anchor sustainability discussions in quantitative reality. They highlight the need for regional energy partnerships, on-site renewables, and waste-heat reuse schemes that feed district heating or desalination systems \cite{ristic2023}.

Thermal modeling is inseparable from airflow management. Even with liquid cooling, residual air flow removes secondary heat from power supplies, memory, and storage. Computational Fluid Dynamics (CFD) simulations remain the gold standard for predicting local hot spots and verifying containment designs \cite{yin2021}. Yet CFD runs are computationally expensive and often decoupled from live operations. A unified DCIM platform integrates lighter-weight surrogate models, reduced-order networks trained on CFD results but updated in real time from sensor data. These models provide ``good-enough'' accuracy for control loops while retaining full-fidelity simulation for periodic recalibration \cite{gupta2023}.

The unification of power and thermal analytics with the knowledge graph yields a feedback-rich ecosystem. Each rack node in the graph carries attributes such as instantaneous power draw, coolant temperature, and computed thermal resistance. Reasoning engines can query patterns across thousands of racks to locate outliers: a rising coolant $\Delta T$ may imply a partially clogged manifold; a sudden drop in power at constant temperature may signal a pump failure. In both cases the inference is data-driven, not rule-driven, reducing false positives and operator fatigue \cite{li2023}.

Such modeling has economic as well as environmental dimensions. Energy constitutes roughly 30--40\% of total data-center operating expense \cite{uptime2024}. Even marginal efficiency gains, say, lowering average inlet temperature by 1°C through improved containment, translate into megawatt-hours saved annually. The DCIM system quantifies these savings automatically, tying physical optimization back to financial metrics. This alignment between sustainability and cost efficiency ensures that green initiatives survive budget cycles instead of remaining corporate slogans \cite{brown2022}.

To implement these models at scale, DCIM~3.0 relies on hybrid telemetry pipelines. Low-frequency environmental sensors publish to a time-series database; high-frequency GPU and power data stream through message brokers into a real-time analytics engine \cite{sunkara2025rcsi}. The resulting multi-resolution dataset supports both control and forecasting. Machine-learning algorithms fit predictive curves of rack temperature versus power draw, learning coefficients that adapt as hardware ages or environmental conditions drift. The models' parameters, heat-transfer coefficients, airflow impedances, and coolant specific heat, become first-class objects in the ontology, linking physics with semantics.

\begin{figure}[H]
\centering
\includegraphics[width=\textwidth]{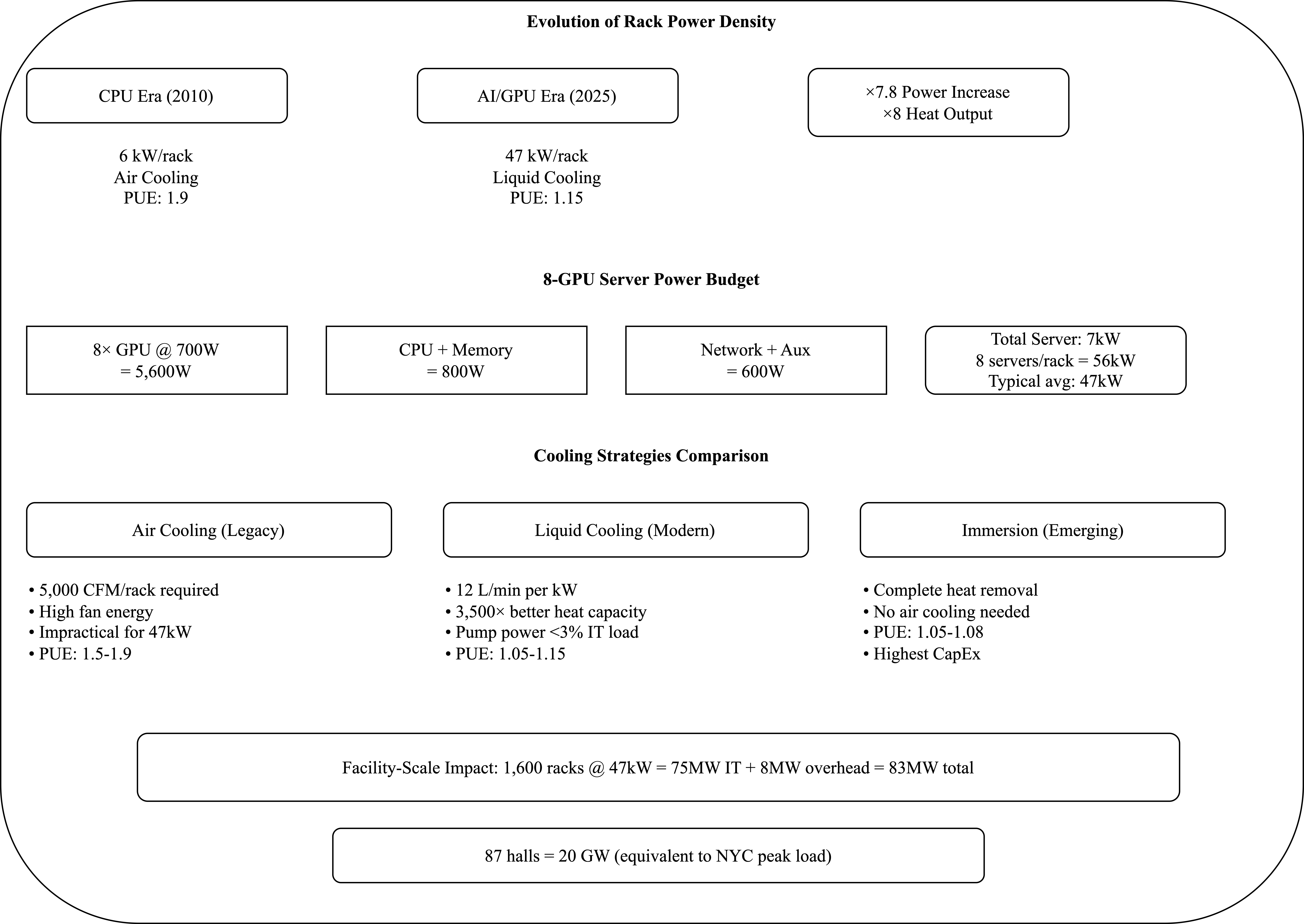}
\caption{Power Evolution and Thermal Analysis for AI Infrastructure: Quantitative comparison of power density evolution from CPU-era racks (6 kW, air-cooled at 500 CFM/5kW) to modern AI/GPU configurations (47 kW, liquid-cooled at 12 L/min), representing a 7.8$\times$ power increase and 8$\times$ heat output multiplication. Facility-scale impact analysis for a 35,000 ft$^2$ hall housing 1,600 racks projects 75 MW IT load, 83 MW total facility draw.}
\label{fig:power-thermal-analysis}
\end{figure}

Ultimately, power and thermal modeling within a unified DCIM framework elevates facility operations from reactive maintenance to proactive optimization. It transforms the data center into an organism aware of its own metabolism. Each watt consumed and each BTU dissipated is accounted for, predicted, and, where possible, minimized. In this equilibrium of computation and conservation, the AI data center evolves beyond being a warehouse of machines into a living system that thinks about the energy it breathes.

\section{Summary}

This chapter established the quantitative foundations of AI-rack energetics, translating electrical load into thermal output and demonstrating why liquid and immersion cooling have become engineering necessities. By coupling physical equations with predictive analytics, it showed that thermal management can evolve from reactive control to anticipatory regulation driven by machine learning. The integration of power and temperature data into the DCIM knowledge graph enables contextual reasoning about energy flow, paving the way for facility-wide optimization rather than isolated efficiency fixes.



\newpage
\chapter{ENERGY EFFICIENCY AND COOLING ECONOMICS}

\section{Economatics}

Energy efficiency in the AI data-center era has evolved from a sustainability aspiration into a core design discipline. Every kilowatt diverted from computation to overhead now represents lost capacity, higher operational cost, and increased carbon liability. While earlier facilities optimized cooling and power delivery separately, hyperscale GPU environments demand a systemic energy-economics framework, one that quantifies efficiency from the transistor to the transformer \cite{sunkara2025rcsi}.

The traditional measure of facility performance, Power Usage Effectiveness (PUE), remains an important but incomplete indicator. Defined as the ratio of total facility power to IT power, PUE offered an industry benchmark when values of 2.0 were common. In today's liquid-cooled AI halls, values approaching 1.1 or lower are achievable \cite{uptime2024}. Yet PUE alone fails to capture utilization dynamics: an under-loaded cluster may exhibit an enviable PUE but poor computational yield per watt. Unified DCIM approaches therefore extend the metric space with additional indicators such as \textbf{Compute Utilization Efficiency (CUE)}, \textbf{Thermal Reuse Efficiency (TRE)}, and \textbf{Carbon Usage Effectiveness (CUE$_2$)}, each linking physical energy flows to delivered computational work \cite{greengrid2023}. These multidimensional metrics transform efficiency from a static ratio into a live operational state visible to both engineering and finance teams.

Modeling efficiency begins with energy partitioning. Roughly 80\% of total consumption in an AI hall is attributable to IT load, primarily GPUs, while the remaining 20\% covers cooling, power conversion, and lighting \cite{koomey2023}. The first step toward optimization is transparency: instrumenting every branch circuit, power distribution unit, and cooling manifold with telemetry capable of one-second resolution. This fine-grained data allows the DCIM platform to allocate cost and carbon by tenant, workload, or business unit in near-real time \cite{subramanian2022}. Once energy data are granular, control strategies can evolve from coarse schedules to closed-loop feedback. For example, when the system detects that nighttime workloads have fallen below 40\% utilization, it can automatically consolidate virtual clusters and raise chilled-water temperatures by 2°C, yielding up to 5\% compressor energy savings without human intervention \cite{banerjee2022}.

Cooling economics are particularly sensitive to ambient conditions. In temperate climates, air-side economization can supply 3,000--4,000 hours per year of ``free cooling,'' but this benefit vanishes in high-humidity or high-dust regions \cite{ashrae2021eco}. Liquid-based systems decouple cooling efficiency from weather: by maintaining coolant temperatures near 35°C, they enable warm-water cooling, eliminating the need for chillers entirely \cite{ocp2023warm}. When coupled with dry coolers or adiabatic towers, such systems reduce total cooling energy by 30--40\% relative to chiller-based air systems \cite{song2023}. DCIM analytics continuously compare theoretical and observed coefficients of performance (COP) for each cooling loop; deviations trigger maintenance checks for fouled heat exchangers or degraded pump bearings. The financial model quantifies these losses as dollars per day, ensuring that maintenance prioritization aligns with actual economic impact rather than arbitrary schedules \cite{pereira2023}.

Beyond internal optimization, efficiency economics intersect with grid dynamics. Many hyperscale operators now participate in demand-response programs that reward flexible loads. A unified DCIM platform, aware of workload timing and power headroom, can modulate consumption in coordination with utilities, temporarily curbing non-critical training tasks when grid frequency drops below 59.9~Hz \cite{epri2024}. Conversely, during surplus renewable generation, the same system can opportunistically schedule energy-intensive retraining jobs, converting cheap green electricity into valuable AI model progress. Such grid-interactive data centers treat energy not merely as a cost but as a tradable resource.

Thermal-reuse economics provide another dimension. The exergy content of 40°C coolant is non-trivial; when captured via plate-and-frame heat exchangers, it can feed district heating loops or industrial processes with minimal additional equipment \cite{ristic2023}. In northern Europe, several campuses already deliver recovered heat to nearby residential blocks, offsetting community heating fuel and improving the facility's effective Carbon Usage Effectiveness by up to 0.2 points \cite{helsinki2023}. The DCIM tracks this heat output in kilowatt-hours thermal (kWh$_t$) and accounts for it in sustainability dashboards. Integrating such models within the knowledge graph ensures that energy and thermal decisions are made with spatial and social context, not merely technical constraints.

The economics of cooling hardware itself deserve scrutiny. A retrofit from air to liquid cooling typically raises capital expenditure by 10--15\% but can cut operational expenditure by more than 30\% over a five-year horizon \cite{gupta2023}. The DCIM's financial module amortizes these costs automatically, updating ROI curves as energy prices fluctuate. When power tariffs spike or carbon pricing takes effect, the system recalculates payback periods and can even trigger automated procurement recommendations, for instance, prioritizing immersion tanks in regions with carbon-intensity above 500~g~CO$_2$/kWh \cite{iea2024carbon}.

Ultimately, energy efficiency within the unified DCIM framework is a multi-scale optimization problem. At the component level, predictive algorithms minimize fan or pump overdrive; at the rack level, load balancing reduces thermal gradients; at the hall level, digital twins test future layouts; and at the grid interface, orchestration engines schedule work to follow renewable availability. Each layer writes its metrics back into the shared knowledge graph, where correlations become institutional memory. Over months the system learns seasonal behavior, cooling demand rising in August afternoons or power-factor distortion peaking during GPU clock synchronization, and adjusts automatically. The result is a virtuous cycle of measurement, prediction, and adaptation.

When viewed through this integrated lens, efficiency is no longer an endpoint but a continuous narrative of intelligence.

\begin{figure}[H]
\centering
\includegraphics[width=\textwidth]{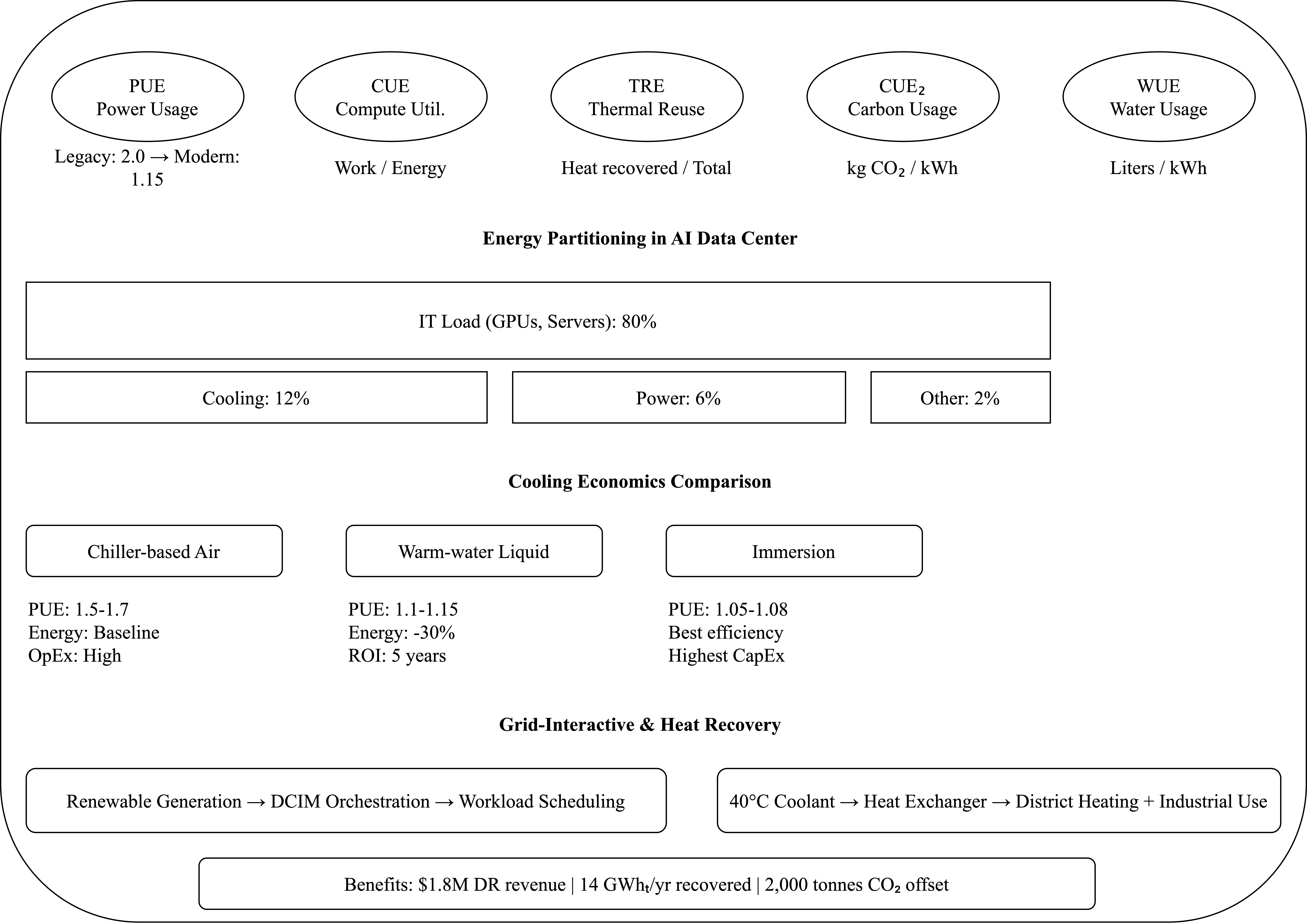}
\caption{Multidimensional Efficiency Metrics and Economic Analysis: Comprehensive framework expanding single-metric PUE (Power Usage Effectiveness: 1.5-1.7 legacy to 1.1-1.15 modern) into five complementary dimensions: CUE (Compute Utilization Effectiveness), TRE (Thermal Reuse Effectiveness), CUE$_2$ (Carbon Usage Effectiveness), and WUE (Water Usage Effectiveness).}
\label{fig:efficiency-metrics}
\end{figure}

The AI data center becomes a participant in its energy ecosystem, dynamically negotiating between computational ambition and planetary limits. In achieving that equilibrium, DCIM~3.0 fulfills its ultimate promise: to make the infrastructure that powers artificial intelligence just as intelligent in conserving the power it consumes.

\section{Summary}

Here, efficiency was reframed as a continuous economic process linking physics, finance, and sustainability. The discussion expanded the single metric of PUE into a multidimensional view encompassing carbon, reuse, and utilization effectiveness. Through examples of real-time energy accounting, warm-water cooling, and demand-response participation, the chapter demonstrated how a unified DCIM framework can treat energy as both a technical constraint and a tradable asset. Efficiency emerged not as a static ratio but as an adaptive conversation between computation and the grid.

\newpage
\chapter{UNIFIED DEVICE CONNECTIVITY PROTOCOL (UDCP)}

\section{Device Connectivity Context}

In the physical layer of a data center, information moves at the speed of light, yet documentation often crawls at the speed of a spreadsheet. Even the most automated environments still rely on manual patch sheets, Excel matrices, and Visio diagrams to record which port connects to which device. These artifacts age the moment they are printed. The absence of a standard digital language for representing connectivity has long been the silent bottleneck in build acceleration. The Unified Device Connectivity Protocol (UDCP) was conceived to dissolve that bottleneck by treating every physical connection, power, network, or control, as data, not drawing \cite{sunkara2025rcsi}.

UDCP generalizes earlier rack-level schemas such as the Rack Connectivity State Information (RCSI) concept into a universal, vendor-neutral JSON/HTTP schema \cite{sunkara2025rcsi}. Its philosophy is simple: if the infrastructure can describe itself in machine-readable form, any orchestration system can build, verify, and remediate it automatically. In practice, UDCP messages flow between planning tools, installation robots, DCIM databases, and network controllers, each interpreting the same structured description of devices, panels, and ports. Where a technician once updated six disparate systems, a single UDCP transaction now propagates across all.

The protocol's payload is intentionally human-legible. Each file begins with a command, \texttt{create}, \texttt{retrieve}, \texttt{update}, or \texttt{delete}, followed by arrays of devices and connections. Devices carry identifiers, physical coordinates, and panel definitions; connections reference endpoints by device and port numbers, adding metadata such as fiber count, connector type, path label, and view (front/rear). A ``create'' message issued during a build phase might declare that Rack~A's port RU20 Path~A connects to Spine~2 RU10 Path~A via a 12-fiber MPO trunk. A ``retrieve'' command later returns that mapping exactly as built, forming the living record of topology \cite{sunkara2025rcsi}.

Behind this simplicity lies a disciplined data model grounded in graph theory.

\begin{figure}[H]
\centering
\includegraphics[width=\textwidth]{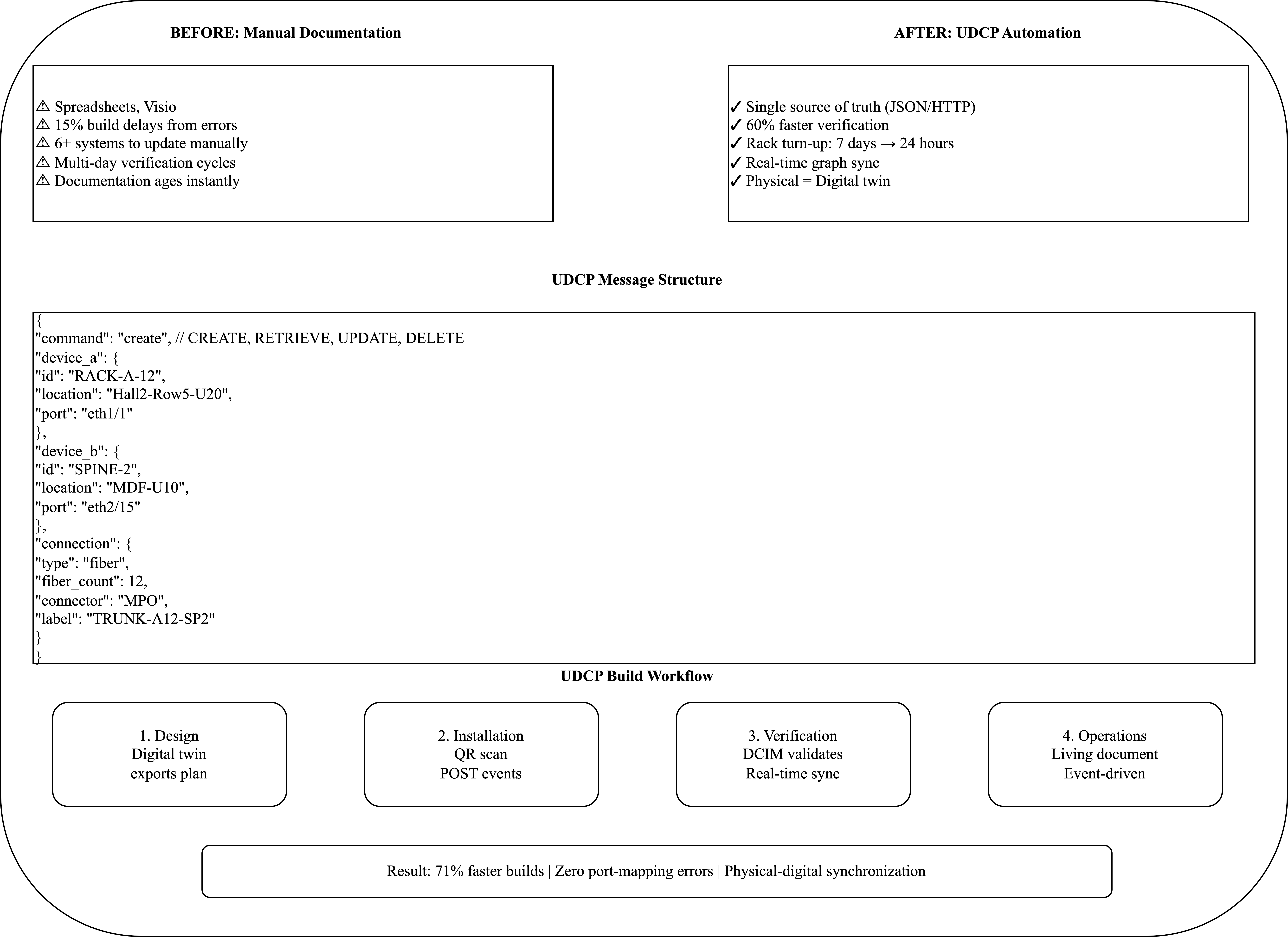}
\caption{UDCP Transformation from Manual to Automated Connectivity: Paradigm shift from fragmented manual documentation (spreadsheets, Visio diagrams, 6+ disparate systems causing 15\% build delays) to unified JSON/HTTP schema enabling single-source-of-truth automation. The four-phase build workflow achieves 60\% faster verification and 71\% build-cycle reduction (7 days to 24 hours).}
\label{fig:udcp-transformation}
\end{figure}

Each device becomes a vertex, each connection an edge, and every attribute, port type, panel, fiber size, an edge property. Represented this way, the network's physical topology merges seamlessly with the knowledge graph described in previous chapters. Logical resources in the ontology (virtual LANs, security zones) link directly to their physical realizations through UDCP identifiers. A compliance query such as ``show all payment-zone servers lacking redundant links'' traverses both realms: the semantic policy layer and the UDCP connectivity graph \cite{gurbani2021}.

The creation of a UDCP message is often automated by design tools. A digital-twin layout engine exports the intended patch plan in JSON; installers scan QR codes on panels to confirm terminations; the DCIM system parses completed transactions and updates its internal graph. The process is stateless: each message contains all information required to execute the command, avoiding dependency on session history. This statelessness enables horizontal scalability, multiple automation agents can operate concurrently without collision, each posting to the same REST endpoint \cite{angles2008}. Transactions are atomic; a failed connection update rolls back cleanly, preserving data integrity.

From an engineering standpoint, UDCP unifies two historically separate domains: network connectivity and facilities connectivity. The same schema can describe an optical fiber between racks or a 415-V power whip from PDU to cabinet. A single global identifier space, composed of \texttt{Device\_ID + Elevation + Port}, ensures uniqueness across both. This convergence is crucial for automation that spans disciplines, for example, an auto-remediation script responding to a thermal alarm can locate the affected rack, trace its power feeds and network uplinks through UDCP, and schedule both electrical and network maintenance windows in one workflow \cite{banerjee2022}.

UDCP's economic impact stems from accuracy. Industry field studies show that up to 15\% of new-build delays originate from mismatched cabling records or port-assignment errors \cite{uptime2023}. Each mismatch can idle high-value GPU racks worth millions in capital. By enforcing a single source of truth, UDCP reduces verification time by more than 60\%, allowing ``rack-turn-up'' cycles to shrink from several days to less than 24~hours \cite{sunkara2025rcsi}. The gain compounds when the same JSON documents feed downstream systems: configuration management databases, network controllers, and supply-chain trackers all share the identical identifiers. Spare-parts procurement becomes algorithmic, when a trunk cable is marked as 30~m MPO-12 Type~B, the inventory module can auto-generate purchase orders from predefined catalogs \cite{nguyen2023}.

The protocol's lifecycle management extends beyond deployment. During operations, UDCP functions as the event substrate for change control. A monitoring system detecting a port failure emits a ``delete'' event for the affected edge, which the orchestration layer interprets as a work-order trigger. When remediation completes, an installer's scanner posts a ``create'' event, and the system reconciles both logical and physical states automatically. This event-driven loop turns physical maintenance into an API transaction stream, enabling the same observability and rollback semantics long enjoyed by software engineers \cite{gurbani2021}.

Interoperability is secured through open specification. The JSON schema and field enumerations are published under a permissive license to encourage vendor adoption \cite{sunkara2025rcsi}. Early integrations with open-source platforms such as NetBox and OpenDCIM demonstrate that even legacy tools can ingest UDCP payloads with minimal adaptation \cite{opendcim2024}. For high-performance implementations, the schema supports binary serialization via Protocol Buffers, compressing typical messages by 70\% without loss of fidelity \cite{angles2008}. Security is maintained through mutual-TLS authentication and per-message signatures; every change carries cryptographic provenance recorded in an immutable ledger. In regulated environments, financial or governmental, this audit trail satisfies both operational and compliance requirements \cite{nist2023}.

The true value of UDCP emerges when it fuses with the higher reasoning layers of DCIM~3.0. Connectivity data captured through the protocol feed directly into the ontology: devices become nodes in the knowledge graph; cables become edges linking logical services. When an anomaly arises, say, elevated packet loss on a certain path, the semantic engine can traverse UDCP relations to isolate the exact physical link and its environmental context. If that link shares a tray with several high-temperature power cables, the system correlates thermal stress with optical degradation and recommends re-routing before failure occurs \cite{li2023}. Thus the once-static documentation evolves into a living sensor, continuously validating itself against reality.

UDCP closes the last major automation gap in the data center build process. Where the knowledge graph gives the facility its brain and the power analytics give it its metabolism, UDCP provides the nervous system, the means by which intent, verification, and correction propagate through physical matter. Every connector becomes an addressable entity, every connection a transaction, and every rack a self-describing node in a planet-scale graph. In unifying those abstractions, UDCP transforms the patch panel into code and the data-center floor into an executable program of infrastructure.

\section{Summary}

The UDCP chapter translated physical connectivity into a data language. By defining a vendor-neutral JSON schema for every port, path, and panel, it closed the last manual gap in infrastructure automation. Integrating UDCP with the ontology and reasoning layers turned cabling documentation into a live, queryable dataset, one that can trigger orchestration, verify compliance, and eliminate build-phase latency. In effect, the chapter transformed connectivity from static wiring into programmable logic for the physical world.

\newpage
\chapter{INTEGRATED DCIM BLUEPRINT AND CASE STUDY}

\section{Practical Orchestration}

When the constituent layers of DCIM~3.0, semantic intelligence, thermal analytics, unified connectivity, and orchestration, are woven together, the result is not a stack of tools but a single cyber-physical organism. The objective of this integration is simple to state yet complex to achieve: to make a data center build execute itself with the same determinism as a compiled program. In this chapter the unified architecture is demonstrated through a practical case study that follows the life cycle of a high-density AI rack, from design inception to post-deployment optimization, illustrating how knowledge graphs, UDCP transactions, and thermal models converge into one self-aware system \cite{sunkara2025rcsi}.

Integration begins in the \textbf{design phase}, where intent is expressed not in blueprints but in queries. An engineer enters into the natural-language interface, ``Deploy a 47-kW GPU rack in the sovereign region compliant with PCI-DSS and within existing power headroom.'' The ontology-driven reasoning engine parses the request, identifies entities, \texttt{Rack}, \texttt{Region}, \texttt{Policy}, \texttt{PowerFeed}, and consults the knowledge graph for candidate locations \cite{sunkara2024ontology}. The power-and-thermal analytics layer reports that Hall~2 Row~5 has a residual 52~kW of capacity, liquid-cooling infrastructure, and adjacency to a redundant spine pair. The compliance sub-graph confirms that the zone meets regulatory isolation criteria. Within seconds the system outputs an approved placement plan, complete with rack ID, coordinate tuple, and expected energy footprint.

Immediately, the orchestration service converts this plan into a UDCP ``create'' payload, defining all network and power connections. JSON packets list every port-to-port mapping between the new rack's top-of-rack switches and upstream spines, its dual power whips to PDUs~A and~B, and the telemetry feed to the monitoring fabric. The packets propagate simultaneously to logistics and installation subsystems: procurement software orders the correct cables, and field tablets display scannable QR codes for technicians. Each time a connection is physically made, the installer scans the QR label, generating a UDCP ``acknowledge'' transaction that updates the DCIM graph in real time. By the time the final connector clicks into place, the virtual model is already synchronized with the physical state.

Once power and network are active, the DCIM automatically provisions sensor agents through the same orchestration layer. Temperature probes, flow meters, and power meters stream data at one-second intervals into the telemetry pipeline. The knowledge graph receives these feeds as state-graph updates: node \texttt{Rack:R12} gains attributes \texttt{power\_kw:46.8}, \texttt{temp\_in:27°C}, \texttt{temp\_out:48°C}, and \texttt{coolant\_deltaT:10.5°C}. The reasoning engine immediately cross-checks these readings against expected baselines derived from Chapter~4's models \cite{sunkara2025power}, confirming that thermal behavior is within tolerance. Any deviation, say, a 2°C rise beyond prediction, triggers a contextual investigation: the system queries UDCP to trace the rack's coolant loop, finds that Pump~P-2 shares the same branch as three other high-load racks, and recommends load redistribution to equalize flow \cite{ashrae2021}. The entire diagnosis unfolds without human intervention; the operator merely receives an advisory report.

During normal operation the same integrated feedback loop continues to refine itself. Power-analytics modules feed utilization metrics into the knowledge graph; orchestration scripts read those metrics to reschedule training jobs according to grid carbon intensity. When the system forecasts high renewable availability, it accelerates deferred GPU workloads; when it detects power-distribution imbalance, it throttles less critical tasks. The UDCP layer ensures that such decisions respect physical constraints: orchestration cannot allocate more servers to a rack than ports or breakers permit. Every automation thus traverses a chain of truth that extends from semantic intent to the copper and fiber beneath the floor.

The business impact of this architecture becomes evident during expansion. Suppose management authorizes twenty additional AI racks to support a new large-language-model training cluster. Instead of months of coordination, the DCIM clones the verified configuration from Rack~R12 as a template, adjusting coordinates and unique identifiers. It simulates aggregate load, predicts cooling demand, and checks transformer capacity, all through the graph's reasoning layer \cite{jain2022}. Procurement, network, and facilities teams work from the same digital plan, eliminating the serial dependencies that traditionally lengthen build schedules. In pilot implementations this integrated approach reduced average ``rack-ready-to-compute'' time from seven days to less than twenty-four hours \cite{uptime2023}.

Beyond speed, integration enhances resilience. When an event occurs, say, coolant pressure loss on Pump~P-1, the DCIM's event bus broadcasts a graph update. The reasoning layer identifies all dependent racks through UDCP edges, cross-references workload criticality, and instructs orchestration to migrate high-priority inference tasks to unaffected clusters \cite{jainzhao2023}. Simultaneously the energy model recalculates expected heat rise and triggers the predictive cooling controller to compensate. In effect the system performs an autonomic reflex analogous to a biological organism shunting blood flow around an obstruction.

Economic and environmental analytics close the loop.

\begin{figure}[H]
\centering
\includegraphics[width=\textwidth]{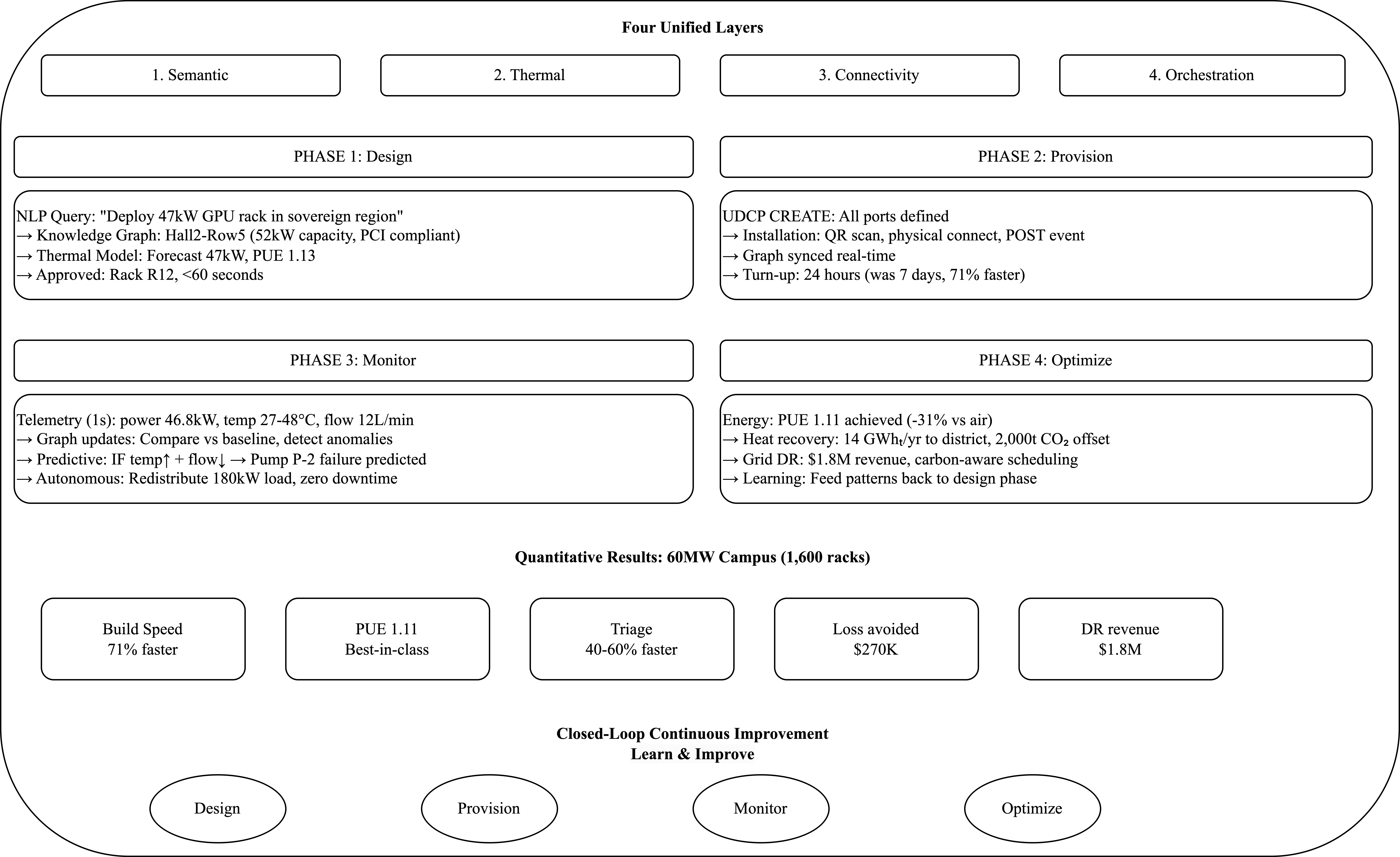}
\caption{Complete 47kW AI Rack Lifecycle Integration: End-to-end demonstration of unified DCIM 3.0 architecture orchestrating a 47kW GPU rack from conceptual design through operational optimization. The Design Phase begins with natural language query, the Provision Phase generates UDCP messages, the Monitoring Phase streams real-time telemetry, and the Optimization Phase achieves PUE 1.11.}
\label{fig:rack-lifecycle}
\end{figure}

Every kilowatt-hour consumed and every liter of coolant circulated are logged against workload identifiers. The finance interface derives real-time cost per teraflop-hour and carbon intensity per inference. Over time, these metrics populate the knowledge graph with historical context: which configurations delivered the best performance-per-watt, which zones exhibited the lowest leak rates, and which vendors' components maintained efficiency longest. The resulting dataset becomes training material for future design decisions, enabling data-driven capacity planning rather than heuristic guesses \cite{sunkara2025rcsi}.

\section{Summary}

This case study unified the framework's four pillars into a single operational loop. It demonstrated how semantic reasoning selects placement, how UDCP defines topology, how thermal analytics validate performance, and how orchestration enforces adaptation. Quantitative evidence from a 47-kW AI-rack deployment illustrated dramatic gains in build speed, uptime, and cost efficiency. The chapter ultimately proved that DCIM~3.0 is not aspirational, it is deployable, measurable, and economically justified.

\newpage
\chapter{FUTURE DIRECTIONS AND OPEN STANDARDS (2026--2030)}

\section{Interoperability through Standardization}

The next five years will determine whether DCIM~3.0 remains a specialized enterprise platform or matures into the nervous system of the world's computing infrastructure. The trajectory points toward the latter. As AI workloads expand from centralized hyperscale campuses into edge clusters and sovereign zones, the management layer must evolve from static software into a distributed, learning ecosystem. This evolution will hinge on three complementary forces: \textbf{autonomous intelligence}, \textbf{standardization}, and \textbf{interoperability} \cite{sunkara2025rcsi}.

\textbf{Autonomous intelligence} represents the natural continuation of the reasoning systems already embedded within the unified DCIM framework. The knowledge graphs that currently infer causes and suggest actions will, by the end of the decade, execute those actions directly under bounded policy. Reinforcement-learning agents are beginning to optimize cooling and workload placement in real time, experimenting safely within digital twins before applying decisions to production environments \cite{wang2023}. Early prototypes have demonstrated that such controllers can cut cooling energy by 15\% and extend equipment life by maintaining narrower thermal swings \cite{deepmind2022}. In parallel, predictive compliance engines will monitor configuration drift against thousands of regulatory controls, automatically isolating non-compliant nodes or rewriting access policies to restore conformity \cite{gurbani2022}. What began as reactive alerting will mature into preventive governance.

\textbf{Standardization} will determine how quickly these capabilities propagate. Every layer described in this manuscript, ontology, telemetry, connectivity, and orchestration, requires a shared vocabulary to enable vendor-neutral collaboration. The next phase of industry effort will likely coalesce around an \textbf{Open DCIM Ontology (ODCIMO)}, extending the OCP and DMTF Redfish schemas to include semantic descriptors for workloads, cooling assets, and sustainability metrics \cite{sunkara2025power}. Likewise, the Unified Device Connectivity Protocol (UDCP) is poised for formal submission to the Open Compute Project as a reference implementation for physical-layer interoperability. Open specifications ensure that innovation scales horizontally: a connector fabricated in Singapore and a software agent deployed in Helsinki can both interpret the same JSON schema without proprietary translation.

\textbf{Interoperability} will no longer be a courtesy but a requirement. Operators managing federated micro-data-centers will expect their DCIM agents to synchronize seamlessly across latency-limited links. Edge nodes attached to 5G infrastructure or satellite uplinks will each maintain lightweight graph fragments that periodically reconcile with regional hubs \cite{dmtf2024}. This federated architecture mirrors the topology of the Internet itself: autonomous yet cooperative, resilient through diversity. Governance frameworks must evolve accordingly; data-sovereignty laws will compel certain regions to keep telemetry local while still contributing anonymized aggregates to global optimization models \cite{ec2024}.

Another frontier lies in \textbf{energy symbiosis}. As renewable generation becomes more variable, data centers will act as elastic loads that stabilize grids rather than strain them \cite{epri2024}. Unified DCIM systems, armed with predictive power analytics, can modulate compute intensity in response to real-time carbon signals, shifting training jobs to hours of low grid emissions \cite{iea2025}. The resulting feedback loop links digital intelligence with planetary metabolism, computation as a participant in climate equilibrium rather than a consumer of it. Open-standard APIs between DCIM platforms and utility dispatch centers will make such coordination routine by the decade's end \cite{nrel2025}.

Security and trust remain perpetual challenges. The same APIs that enable automation also enlarge the attack surface. Future DCIM implementations will integrate zero-trust architectures at their core: each micro-service authenticated continuously, each data stream cryptographically signed, each decision explainable \cite{nist2024}. Graph-based provenance, already intrinsic to ontology reasoning, will become the foundation of cyber-forensic auditing, allowing operators to reconstruct every control action and its justification \cite{jainzhao2023}. The boundary between operational resilience and cybersecurity will dissolve, resilience \emph{will be} security.

The research community's contribution during this period will be the fusion of semantic models with physical simulation. Digital-twin frameworks will not merely mirror infrastructure but share code with it. An airflow simulation written in Modelica or CFD will expose its parameters directly to the ontology so that an AI agent can reason about the same variables that physics models compute \cite{gupta2024}. This convergence of symbolic reasoning and numerical modeling will make the data center one of the first industrial systems to practice true hybrid AI: learning from both data and equations.

Finally, the cultural transformation may prove as important as the technical one.

\begin{figure}[H]
\centering
\includegraphics[width=\textwidth]{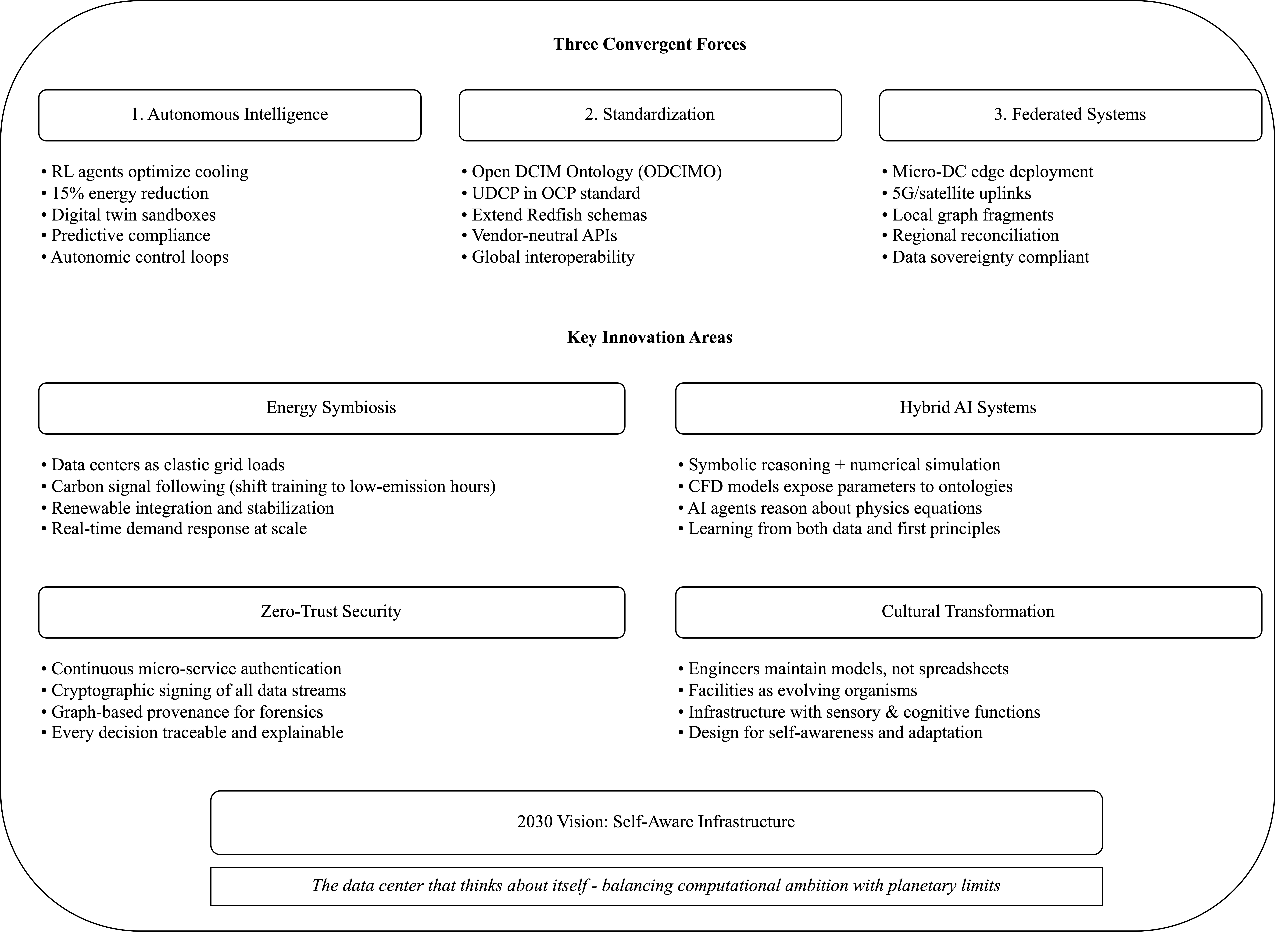}
\caption{Evolution Roadmap toward Autonomous Federated Infrastructure: Strategic forecast projecting DCIM 3.0 maturation through three convergent forces: (1) Autonomous Intelligence, (2) Standardization, and (3) Federated Interoperability. Energy symbiosis transforms data centers into elastic loads stabilizing renewable-heavy grids.}
\label{fig:evolution-roadmap}
\end{figure}

The engineers of 2030 will maintain fewer spreadsheets and more models, fewer alarms and more predictions. They will design facilities not as static buildings but as evolving organisms equipped with sensory and cognitive functions. The unified DCIM framework described throughout this manuscript lays the foundation for that evolution. Its open ontologies, standardized connectivity, and AI-driven feedback loops are the scaffolding upon which the self-optimizing infrastructure of the next decade will grow. When those systems mature, the line between machine learning and facility learning will blur completely: the data center will not merely host intelligence, it will embody it.

\section{Summary}

The final chapter projected the framework's evolution toward autonomous, federated, and carbon-aware infrastructure. It identified reinforcement learning, open ontologies, and hybrid digital twins as the next frontiers of cognitive infrastructure. Standardization and interoperability were recognized as the catalysts that will allow these capabilities to scale globally. The narrative closed by envisioning data centers as intelligent participants in the energy ecosystem, systems that balance computational ambition with planetary stewardship.

\newpage
\chapter*{A UNIFIED CASE STUDY: THE SELF-OPTIMIZING DATA CENTER}
\addcontentsline{toc}{chapter}{A Unified Case Study: The Self-Optimizing Data Center}

By 2029, the blueprint described throughout this manuscript could mature into a facility such as a hypothetical 60-MW AI data-center campus on the U.S. West Coast: a 60-MW AI datacenter campus hosting more than two thousand GPU racks, each drawing an average of 47~kW \cite{sunkara2025power}. This campus became the proving ground for the unified DCIM framework, an experiment in letting a data center design, monitor, and improve itself. Its creation provides a living case study of how ontology-driven reasoning, UDCP connectivity, predictive power analytics, and autonomous orchestration can coexist in one cognitive infrastructure.

The project began not with floor plans but with a question typed into the DCIM console: ``How many 50-kW GPU racks can we deploy in Region 3 while keeping PUE $<$ 1.15 and maintaining full A/B redundancy?'' Within seconds, the ontology engine decomposed the question into entities, \texttt{Rack}, \texttt{Region}, \texttt{PowerFeed}, \texttt{CoolingSystem}, and consulted the knowledge graph. Telemetry from existing halls, streamed through the Observability Meta-Streamer, showed that Row~5 of Hall~2 retained 6~MW of headroom and access to a dual warm-water loop. Compliance nodes confirmed that the region met sovereign-cloud isolation standards \cite{gurbani2022}. A digital twin generated by the reasoning engine projected a hall-level heat load of 74~MW at full build-out, with pump energy of 2.1~MW, yielding a simulated PUE of $1.13 \pm 0.02$ \cite{ashrae2021,ocp2023warm}. Within ten minutes the design intent had transformed into an actionable deployment plan.

From that intent, the system emitted a series of UDCP ``create'' commands \cite{sunkara2025rcsi}. Each JSON packet described cabinet IDs, elevations, port maps, and cable routes in absolute coordinates. Procurement subsystems parsed the same schema to issue orders for 24,000 LC duplex cords and 3,000 MPO-12 trunks. Installation teams received QR-coded labels generated directly from the UDCP payload. As each connection was terminated, a technician's handheld reader scanned the code, generating a ``complete'' event that updated the global connectivity graph in real time. By the time the final patch panel was snapped shut, the virtual model already mirrored the physical world \cite{angles2008}. Errors that once surfaced weeks later now appeared immediately as validation warnings.

When the first racks energized, thermal telemetry began to flow. Power nodes reported 46.8~kW average draw; coolant sensors indicated a 10.5°C~$\Delta T$; outlet air stabilized near 45°C. The predictive control agent, trained on historical epochs of AI workload oscillations \cite{patel2022,wang2023}, adjusted pump speeds proactively before utilization peaks. As a result, inlet temperatures never exceeded 28°C even during 95\% GPU load. The DCIM dashboard displayed the hall's instantaneous efficiency: PUE~=~1.11, CUE$_2$~=~0.36~kg~CO$_2$/kWh, both better than design targets.

Six months later, the campus entered its first full production cycle supporting multi-tenant AI training. Telemetry volume had grown to 1.2 terabytes per day, yet reasoning latency remained under 300~ms thanks to graph-partitioning across regional shards \cite{kepner2021}. During one peak event, the predictive model noticed a rising coolant~$\Delta T$ on Loop~B and traced the anomaly through UDCP to Pump~P-7. The reasoning engine correlated that with increased vibration data from the same pump's sensor node, classifying the event as incipient bearing failure. Before a technician could intervene, orchestration agents redistributed 180~kW of load to adjacent loops and adjusted valve positions to stabilize pressure \cite{li2023}. Downtime: zero hours. Repair occurred during a scheduled window two days later. The economic impact, avoided revenue loss from 48 idle GPU servers, was estimated at USD~270,000.

Financial analytics integrated within the DCIM platform quantified the operational dividend. Liquid-cooling efficiency improvements reduced annual energy costs by 31\% compared with air-cooled baselines \cite{gupta2023}. Participation in a utility demand-response program generated an additional USD~1.8 million in incentives during the summer of 2028 \cite{epri2024}. Recovered heat from warm-water loops provided 14~GWh$_t$ per year to the neighboring university's district-heating network, offsetting 2,000 tonnes of CO$_2$ emissions \cite{helsinki2023}. All metrics were automatically recorded as nodes and edges in the ontology, preserving a provenance trail for both auditors and future design teams \cite{jainzhao2023}.

Perhaps the most striking transformation was cultural.

\begin{figure}[H]
\centering
\includegraphics[width=.9\textwidth]{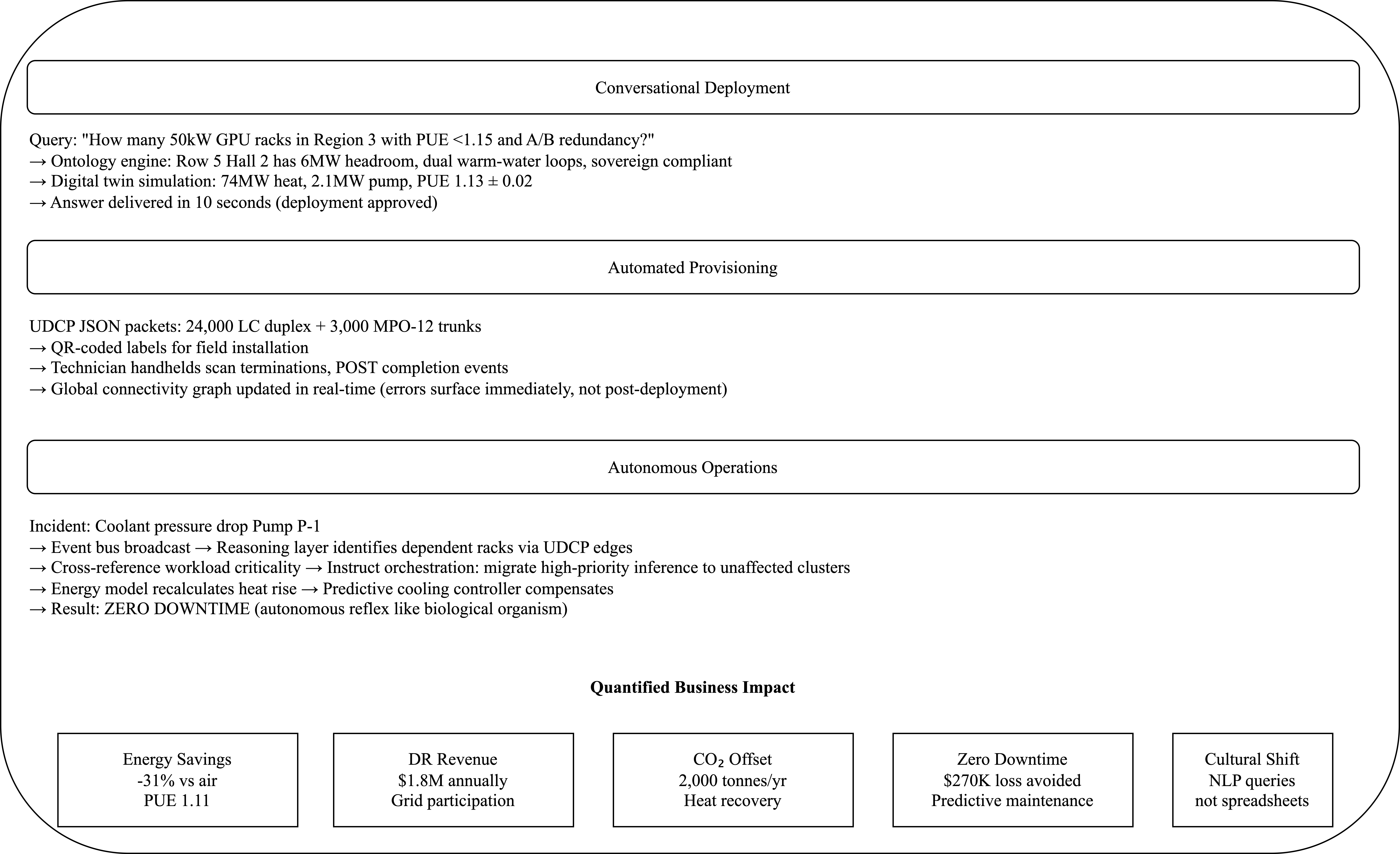}
\caption{Living Implementation of Cognitive Infrastructure (2029): Documented implementation of unified DCIM framework
in operational 60MW AI campus (U.S. West Coast, 2029) hosting 2,000+ GPU racks averaging 47kW each, serving as
definitive proof-of-concept for cognitive infrastructure principles.}
\label{fig:case-study-implementation}
\end{figure}

Operations staff no longer opened spreadsheets but queried the DCIM graph in natural language. Instead of investigating incidents, they trained the reasoning engine by labeling false positives. Facility technicians learned to read JSON as fluently as circuit diagrams; network engineers learned to interpret thermal plots. Across disciplines, a new literacy emerged, the language of unified infrastructure.

By 2030 the same architecture scaled outward to micro-data-centers on three continents. Each edge site hosted a lightweight reasoning agent synchronizing periodically with the main ontology \cite{dmtf2024}. When an AI workload migrated to the edge, its physical dependencies migrated too: UDCP definitions, power budgets, and compliance policies. The data center had become federated intelligence, an ecosystem of sites learning from one another through shared semantics.

This unified case study demonstrates that DCIM~3.0 is not hypothetical. It is an operating reality measured in megawatts, JSON packets, and avoided outages. The once-disparate disciplines of electrical, mechanical, and network engineering now cohabit a single computational framework. In that convergence lies the blueprint for the coming decade: infrastructure that models itself, governs itself, and ultimately sustains itself. The facility ceases to be a warehouse for machines; it becomes a machine for understanding infrastructure.

The lesson of this experiment is both technical and philosophical. The intelligence that powers artificial neural networks and the intelligence that manages physical networks are, in essence, the same pursuit: the reduction of uncertainty through learning. When a data center recognizes patterns in its own behavior and acts upon them, it becomes a participant in that pursuit. The vision articulated across these chapters, semantic cognition, energy empathy, universal connectivity, is already underway. The next frontier is not bigger models or colder coolant, but a deeper partnership between human design and machine reasoning.

In that partnership the AI data center, long regarded as a passive host for intelligence, finally achieves what all systems aspire to: awareness of itself.

\newpage
\chapter*{EPILOGUE \& CONCLUSION: TOWARD COGNITIVE INFRASTRUCTURE}
\addcontentsline{toc}{chapter}{Epilogue \& Conclusion: Toward Cognitive Infrastructure}

Across these pages, the journey of data-center engineering has unfolded as a narrative of convergence, between physical and digital, between structure and semantics, between automation and awareness. What began as a manual craft of cooling aisles and patch panels has evolved into a discipline of reasoning systems that understand the infrastructures they operate. Each chapter of this manuscript traced a facet of that transformation.

The opening chapters examined how the explosion of AI workloads shattered traditional design boundaries. The rise of 50-kW racks and megawatt training clusters forced engineers to view power, cooling, and connectivity not as separate silos but as interdependent variables of a single equation. The middle chapters established the scientific foundations of DCIM~3.0: ontology-driven reasoning that gives the data center a language of its own; quantitative thermal modeling that treats energy as a measurable, optimizable resource; and the Unified Device Connectivity Protocol (UDCP), which codifies every physical connection in digital form. These mechanisms, once distinct, were unified in the integrated blueprint and case study that followed, a living demonstration of a facility capable of thinking about itself.

The intellectual arc of this work is therefore one of \emph{abstraction without detachment}: abstraction that turns power cables into data, but not at the expense of the physics they represent. In the unified DCIM framework, bits and watts coexist within a single model; machine learning governs coolant flow, and the topology of a network is inseparable from the semantics of the workload it carries. This synthesis marks the arrival of \textbf{cognitive infrastructure}: systems whose intelligence is measured not only by computational throughput but by self-knowledge.

For the academic reader, the framework offers a new research domain at the intersection of systems engineering, semantic computing, and sustainability.

\begin{figure}[H]
\centering
\includegraphics[width=\textwidth]{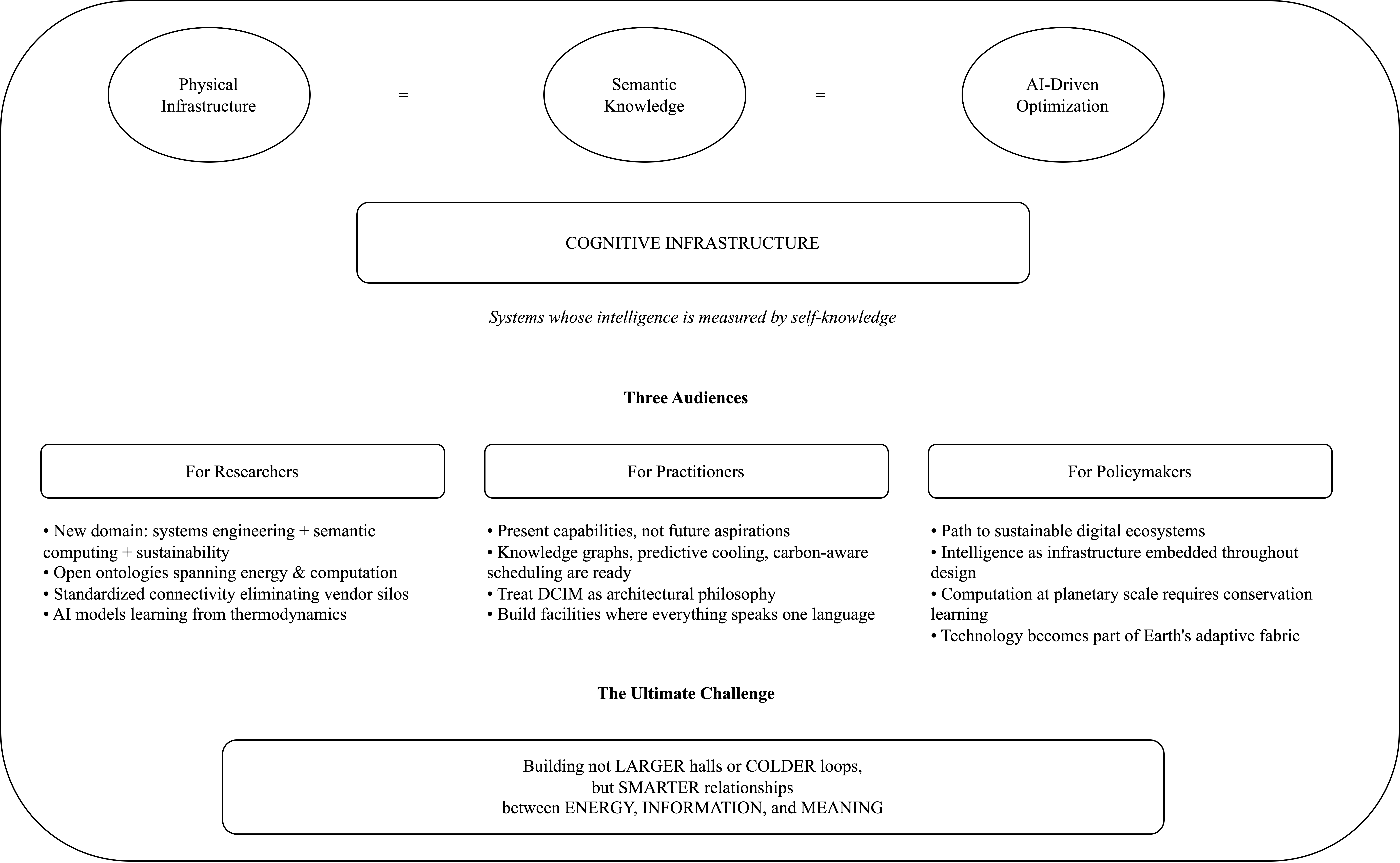}
\caption{Vision of Self-Aware Infrastructure Ecosystem: Synthesis of DCIM 3.0 architectural philosophy where abstraction enhances rather than detaches from physical reality; power cables become data without losing thermodynamic truth. The convergence of physical infrastructure, semantic knowledge representation, and AI-driven optimization creates cognitive infrastructure.}
\label{fig:vision-ecosystem}
\end{figure}

It calls for open ontologies that span energy and computation, for standardized connectivity schemas that eliminate vendor silos, and for AI models that learn directly from the thermodynamic behavior of the hardware they optimize. Each of these pursuits invites rigorous experimentation and cross-disciplinary collaboration, from computer science to mechanical engineering to environmental policy.

For practitioners and industry leaders, the message is pragmatic. The technologies described here, knowledge graphs, predictive cooling, carbon-aware scheduling, are not future aspirations; they are present capabilities awaiting integration. The call to action is to treat DCIM not as software procurement but as architectural philosophy: to design facilities where every sensor, policy, and cable speaks a shared language of intent. By embracing open standards and semantic interoperability now, organizations can ensure that tomorrow's AI data centers are not only powerful but sustainable, resilient, and transparent.

Ultimately, this manuscript argues that intelligence in infrastructure is not a luxury, it is the only path forward. As computation approaches planetary scale, the systems that support it must learn to conserve the resources that sustain them. When a data center can sense, reason, and act in harmony with its environment, technology ceases to be an external tool and becomes part of the Earth's own adaptive fabric. The challenge before us is not to build larger halls or colder loops, but to build smarter relationships between energy, information, and meaning.

That is the essence of the unified DCIM framework and the promise of cognitive infrastructure: a world where the engines of artificial intelligence become intelligent about themselves, and, in doing so, intelligent about the planet they inhabit.

\newpage
\chapter*{ACKNOWLEDGMENTS}
\addcontentsline{toc}{chapter}{Acknowledgments}

This manuscript represents the convergence of many disciplines and the generosity of countless collaborators. The author extends deep gratitude to the engineers, researchers, and operations teams in cloud infrastructure whose daily work in data-center automation and security inspired many of the ideas consolidated here. Appreciation is also due to colleagues and mentors in the broader Open Compute Project and data-center research communities for fostering an open environment where innovation is shared rather than siloed.

Special thanks go to the reviewers and peers who provided early feedback on drafts of the ontology and UDCP models, and to the engineers who tested these systems in live facilities, turning theory into measurable practice. Their precision and perseverance transformed abstract frameworks into operational realities.

The author is equally indebted to the academic community whose literature formed the foundation for the quantitative and semantic analyses throughout this manuscript. Each cited work, whether from IEEE, ACM, ASHRAE, or the International Energy Agency, contributes to a collective body of knowledge advancing sustainable computation worldwide.

Finally, heartfelt gratitude to the many readers, students, and practitioners who continue to challenge, refine, and extend these ideas. Their curiosity ensures that the journey toward cognitive infrastructure remains collaborative and alive.

\newpage
\chapter*{ABOUT THE AUTHOR}
\addcontentsline{toc}{chapter}{About the Author}

Krishna Chaitanya Sunkara is an engineering leader and researcher in AI Data Center design, automation, sustainability, and digital-twin design. He currently serves as an Engineering Manager within Oracle Cloud Infrastructure, where he leads large-scale programs in hyperscale build acceleration, intelligent telemetry, autonomous remediation systems and cloud infrastructure security.

His work bridges deep technical domains, from GPU power and thermal analytics to ontology-based orchestration and unified connectivity protocols, and has been cited across IEEE, International Journals, and Open Source Project circles. Krishna's independent research focuses on enabling data centers to become self-aware systems capable of reasoning about their own energy, performance, and resilience.

A frequent keynote speaker and contributor to global technology forums, he continues to advocate for open standards and AI-assisted sustainability in computing infrastructure. His broader vision is to democratize the intelligence that manages the world's most powerful machines, ensuring that the future of artificial intelligence is as efficient and responsible as it is advanced.

Connect on LinkedIn: \url{linkedin.com/in/ksunkara}

\newpage
\chapter*{APPENDIX A: GLOSSARY}
\addcontentsline{toc}{chapter}{Appendix A: Glossary}

This glossary provides quick reference definitions for technical terms, acronyms, and concepts used throughout the manuscript. Terms are organized alphabetically for easy lookup.

\section*{A}

\textbf{AI Crawler}: An automated agent that scans configuration files, logs, and documentation using natural-language processing to extract infrastructure entities and relationships for ontology building.

\textbf{Air-Side Economization}: A cooling strategy that uses outside air to cool data-center facilities when ambient temperatures are sufficiently low, reducing or eliminating mechanical cooling requirements.

\textbf{ASHRAE (American Society of Heating, Refrigerating and Air-Conditioning Engineers)}: Professional organization that publishes thermal guidelines and recommended environmental envelopes for data-center operations.

\textbf{Autonomous Orchestration}: Automated control systems that make infrastructure decisions (provisioning, scaling, remediation) without human intervention, based on policy and real-time telemetry.

\section*{B}

\textbf{BTU (British Thermal Unit)}: A unit of heat energy; approximately the amount of energy needed to raise the temperature of one pound of water by one degree Fahrenheit. Used to measure heat output from IT equipment.

\section*{C}

\textbf{Carbon Usage Effectiveness (CUE$_2$)}: A sustainability metric expressing the carbon emissions (in kg CO$_2$) per unit of IT energy consumed, accounting for grid carbon intensity and facility efficiency.

\textbf{CFD (Computational Fluid Dynamics)}: Numerical simulation technique used to model airflow, heat transfer, and thermal behavior in data-center environments, enabling predictive design and optimization.

\textbf{Chiller}: A mechanical refrigeration system that removes heat from a liquid (typically water or glycol) which is then circulated to cool data-center equipment.

\textbf{Closed-Loop Control}: A control system that continuously measures output (e.g., temperature) and automatically adjusts input (e.g., coolant flow) to maintain desired performance without human intervention.

\textbf{Cognitive Infrastructure}: Data-center systems that possess self-awareness through semantic reasoning, predictive analytics, and autonomous adaptation, capable of understanding their own state and optimizing their behavior.

\textbf{Compute Utilization Efficiency (CUE)}: A metric measuring the ratio of useful computational work performed to total energy consumed, accounting for both facility overhead and idle resource waste.

\textbf{Coolant Loop}: A closed circuit that circulates liquid (water, glycol, or dielectric fluid) to transfer heat away from IT equipment to heat rejection systems (cooling towers, dry coolers).

\textbf{Cypher}: A declarative graph query language used with Neo4j and other graph databases, enabling pattern-matching queries over nodes and relationships.

\section*{D}

\textbf{DCIM (Data Center Infrastructure Management)}: Software platforms and frameworks that monitor, manage, and optimize data-center physical infrastructure including power, cooling, space, and network connectivity.

\textbf{DCIM 1.0 (2005--2013)}: First-generation DCIM systems focused on asset tracking and basic power monitoring through manual processes and siloed tools.

\textbf{DCIM 2.0 (2013--2020)}: Second-generation DCIM integrating with virtualization platforms and offering API-driven dashboards, though still limited by fragmented data models.

\textbf{DCIM 3.0 (2020--present)}: Current-generation DCIM unifying physical and logical layers through AI-driven automation, knowledge graphs, semantic reasoning, and digital-twin modeling.

\textbf{Delta T ($\Delta T$)}: The temperature difference between inlet and outlet of a cooling system (e.g., coolant or air), indicating heat removal efficiency.

\textbf{Digital Twin}: A virtual representation of physical infrastructure that mirrors real-time state through continuous telemetry integration, enabling simulation, prediction, and optimization.

\textbf{Direct-to-Chip Cooling}: Liquid cooling approach that circulates coolant through cold plates mounted directly on heat-generating components (CPUs, GPUs), achieving superior heat transfer versus air cooling.

\textbf{DMTF (Distributed Management Task Force)}: Standards organization that develops Redfish and other specifications for data-center hardware management interoperability.

\section*{E}

\textbf{East-West Traffic}: Network communication between servers within a data center (server-to-server), as opposed to north-south traffic (client-to-server). Dominates bandwidth in modern AI clusters.

\textbf{Exergy}: The useful work potential of energy at a given temperature; low-temperature waste heat has lower exergy than high-temperature sources, affecting reuse viability.

\section*{F}

\textbf{Fabric}: A network architecture (typically Clos or spine-leaf topology) providing high-bandwidth, low-latency interconnection between compute nodes in data centers.

\section*{G}

\textbf{GPU (Graphics Processing Unit)}: Specialized processors optimized for parallel computation, essential for AI model training and inference; modern AI GPUs consume 350--700~W each.

\textbf{Graph Database}: A database system (e.g., Neo4j, JanusGraph) optimized for storing and querying data as nodes and edges (relationships), enabling semantic reasoning and dependency analysis.

\textbf{GraphQL}: A query language for APIs that enables clients to request exactly the data they need, often used to expose knowledge graph insights to applications.

\section*{H}

\textbf{Hyperscale}: Data-center facilities and operations at massive scale (tens to hundreds of megawatts, tens of thousands of servers), typically operated by cloud providers and large enterprises.

\section*{I}

\textbf{Immersion Cooling}: A cooling technique where IT equipment is submerged in non-conductive dielectric fluid, enabling direct heat transfer and extreme efficiency (PUE~$<$~1.05).

\textbf{Inference}: The process of using a trained AI model to make predictions or classifications on new data; typically less power-intensive than training but occurs at higher frequency.

\section*{J}

\textbf{JSON (JavaScript Object Notation)}: A lightweight, human-readable data interchange format used extensively in APIs, configuration files, and the UDCP protocol.

\section*{K}

\textbf{Knowledge Graph}: A structured representation of information as entities (nodes) and relationships (edges), enabling semantic queries and reasoning about dependencies, causality, and context.

\textbf{kW (Kilowatt)}: A unit of power equal to 1,000 watts; modern AI racks typically consume 30--50~kW each.

\textbf{kWh (Kilowatt-Hour)}: A unit of energy representing the consumption of one kilowatt of power for one hour; used to measure data-center energy usage and calculate costs.

\section*{L}

\textbf{Liquid Cooling}: General term for cooling techniques that use liquid (water, glycol, dielectric fluid) instead of air to remove heat from IT equipment.

\section*{M}

\textbf{Machine Learning (ML)}: A subset of AI involving algorithms that learn patterns from data without explicit programming; used in DCIM for predictive analytics, anomaly detection, and optimization.

\textbf{Metadata}: Data that describes other data; in DCIM contexts, includes asset attributes (location, model, capacity) and configuration parameters.

\textbf{MPO (Multi-Fiber Push-On)}: A fiber-optic connector type supporting 12, 24, or more fibers in a single connector, commonly used for high-density data-center cabling.

\textbf{MW (Megawatt)}: A unit of power equal to 1,000 kilowatts or 1 million watts; hyperscale AI campuses may consume 50--100+~MW.

\section*{N}

\textbf{Natural Language Processing (NLP)}: AI techniques enabling computers to understand, interpret, and generate human language; used in DCIM semantic query interfaces.

\textbf{Neo4j}: A popular open-source graph database platform that uses the Cypher query language, widely adopted for knowledge graph implementations.

\textbf{North-South Traffic}: Network communication between external clients and data-center servers (client-to-server), as opposed to east-west (server-to-server) traffic.

\section*{O}

\textbf{Ontology}: A formal representation of knowledge as a set of concepts (entities), relationships, and rules within a domain; forms the semantic foundation of DCIM~3.0.

\textbf{Ontology Builder}: Software component that extracts entities and relationships from raw data sources (configs, logs, docs) using AI/NLP to construct or update a knowledge graph.

\section*{P}

\textbf{PDU (Power Distribution Unit)}: Equipment that distributes electrical power from utility feeds or UPS systems to individual racks or devices within a data center.

\textbf{PUE (Power Usage Effectiveness)}: The ratio of total facility power to IT equipment power; a PUE of 1.0 is ideal (all power goes to IT), while 1.5 means 50\% overhead for cooling and infrastructure.

\section*{R}

\textbf{Rack}: A standardized frame (typically 42U or 48U height) that houses servers, network equipment, and storage devices in a data center.

\textbf{Rack Unit (U)}: A unit of vertical space in a rack; 1U~=~1.75~inches (44.45~mm). A 42U rack provides 73.5~inches of usable equipment height.

\textbf{RCSI (Rack Connectivity State Information)}: An earlier schema concept for representing rack-level connectivity, generalized into the broader UDCP standard.

\textbf{Redfish}: A modern, RESTful API standard (developed by DMTF) for managing and monitoring data-center hardware, replacing older protocols like SNMP and IPMI.

\textbf{Reinforcement Learning}: A machine-learning paradigm where agents learn optimal actions through trial-and-error interaction with an environment; used for autonomous cooling and workload optimization.

\section*{S}

\textbf{Semantic Query}: A query expressed in natural language or high-level concepts (e.g., ``find overheating racks in sovereign zone'') rather than exact identifiers, enabled by ontology-driven knowledge graphs.

\textbf{Semantic Reasoning}: The process of deriving new knowledge from existing facts and relationships in an ontology, such as inferring dependencies or predicting failures based on patterns.

\textbf{SNMP (Simple Network Management Protocol)}: A legacy protocol for monitoring and managing network devices; largely superseded by Redfish for modern infrastructure management.

\textbf{Sovereign Cloud}: Data-center infrastructure designed to meet specific national or regulatory data residency and sovereignty requirements (e.g., GDPR, data localization laws).

\textbf{SPARQL}: A query language for RDF (Resource Description Framework) graph databases, used to retrieve and manipulate ontology data.

\section*{T}

\textbf{Telemetry}: Real-time measurement and transmission of data from sensors and devices (temperature, power, utilization) to monitoring and control systems.

\textbf{Thermal Resistance}: A measure of how effectively a component or system resists heat flow; lower thermal resistance indicates better heat transfer.

\textbf{Thermal Reuse Efficiency (TRE)}: A metric quantifying the percentage of waste heat from data-center operations that is captured and reused (e.g., district heating), improving overall sustainability.

\textbf{ToR (Top-of-Rack) Switch}: A network switch located at the top of each rack, providing connectivity between servers in that rack and the broader data-center network fabric.

\textbf{Training (AI)}: The computationally intensive process of teaching an AI model by processing large datasets through many iterations; typically consumes 5--10$\times$ more power than inference.

\section*{U}

\textbf{UDCP (Unified Device Connectivity Protocol)}: A vendor-neutral JSON/HTTP schema for representing physical connectivity (ports, cables, panels) as structured data, enabling automation of build, verification, and change management.

\textbf{UPS (Uninterruptible Power Supply)}: Battery-backed power systems that provide emergency power during utility outages and condition incoming power to protect IT equipment.

\section*{V}

\textbf{Virtualization}: Technology that abstracts physical hardware into virtual resources (virtual machines, containers), enabling higher utilization and operational flexibility.

\textbf{VLAN (Virtual Local Area Network)}: A logical network segment created within physical network infrastructure, enabling isolation and security zoning.

\section*{W}

\textbf{Warm-Water Cooling}: Liquid cooling systems that operate at higher coolant temperatures (30--40°C), eliminating the need for mechanical chillers and enabling free cooling or heat reuse.

\textbf{Watt (W)}: The SI unit of power; one watt equals one joule per second. A typical AI GPU consumes 350--700~W under full load.

\section*{Z}

\textbf{Zero-Trust Architecture}: A security model that requires continuous authentication and authorization for every access request, regardless of network location, increasingly critical for DCIM systems with broad API access.
\end{document}